\documentclass[column,preprintnumbers,amsmath,amssymb]{revtex4}
\usepackage{amsmath}
\usepackage{extarrows}
\usepackage[ruled]{algorithm2e}
\usepackage{graphicx}% Include figure files
\usepackage{dcolumn}% Align table columns on decimal point
\usepackage{bm}% bold math
\usepackage[all]{xy}
\usepackage{indentfirst}
\usepackage{amsmath}
\usepackage{multirow}
\usepackage{mathrsfs}
\usepackage{bbm}
\usepackage{euscript}
\usepackage{amssymb}
\begin{document}
%%%%%%%%%%%%%%%%%%%%%%%%%%%%%%%%%%%%%%%%%%%%%%%%%%%%%%%%%%%%%%%%%%%%%%%%%%%%%
%%%%%%%%%%%%%%%%%%%%%%%%%%%%%%%%%%%%%%%%%%%%%%%%%%%%%%%%%%%%%%%%%%%%%%%%%%%%%
\title{Undetectable Werner states using linear Bell inequalities}
%%%%%%%%%%%%%%%%%%%%%%%%%%%%%%%%%%%%%%%%%%%%%%%%%%%%%%%%%%%%%%%%%%%%%%%%%%%%%
 \author{Ming-Xing Luo}

\affiliation{\small $^1$Information Security and National Computing Grid Laboratory,
\\
\small Southwest Jiaotong University, Chengdu 610031, China,
\\
\small $^2$CSNMT, International Cooperation Research Center of China, Chengdu 610031, China,
\\
\small $^3$Department of Physics, University of Michigan, Ann Arbor, MI 48109, USA}

\begin{abstract}
Bell inequality serves as an important method to detect quantum entanglement, a problem which is generally known to be NP-hard. Our goal in this work is to detect Werner states using linear Bell inequality. Surprisingly, we show that Werner states of almost all generalized multipartite Greenberger-Horne-Zeilinger (GHZ) states cannot be detected by homogeneous linear Bell inequalities with dichotomic inputs and outputs of each local observer. The main idea is to estimate the largest violations of Werner states in the case of general linear Bell inequalities. The presented method is then applied to Werner states of all pure states to show a similar undetectable result. Moreover, we provide an accessible method to determine undetectable Werner states for general linear Bell inequality including sub-correlations. The numeric algorithm shows that there are nonzero measures of Werner states with small number of particles that cannot be detected by general linear Bell inequalities.
\end{abstract}
%%%%%%%%%%%%%%%%%%%%%%%%%%%%%%%%%%%%%%%%%%%%%%%%%%%%%%%%%%%%%%%%%%%%%%%%%%%%%
%%%%%%%%%%%%%%%%%%%%%%%%%%%%%%%%%%%%%%%%%%%%%%%%%%%%%%%%%%%%%%%%%%%%%%%%%%%%%
%%%%%%%%%%%%%%%%%%%%%%%%%%%%%%%%%%%%%%%%%%%%%%%%%%%%%%%%%%%%%%%%%%%%%%%%%%%%%
\maketitle

\section*{Introduction}

Einstein-Podolski-Rosen (EPR) believed that local hidden variable (LHV) model is a possible underlying explanation of physical reality \cite{EPR1}. Bell realized that the simple assumptions behind any local hidden variable theory lead to non-trivial restrictions on the strength of correlations and provide an elegant way of verifying the validity of EPR's belief \cite{Bell1}. The idea has been described clearly as the violation of special inequalities which are derived from LHV models. Bell theory \cite{Bell1} has become an utmost importance tool for various applications in quantum information \cite{Bert} going beyond its original motivations. Special Bell inequalities provide evidences for proof of unconditional security for quantum key distribution \cite{Ek,MY,Acin1,Acin2}. They are also useful for experimentally detecting entanglement without priori hypothesis on the behavior of the experiment \cite{Cabel,Brunn,Kimb,Piron}, accomplishing multipartite interactive proof \cite{Bra,Ben,Kemp}, estimating dimensions of Hilbert space \cite{Brun2,Wehn}, for others see reviews \cite{BCMD,BCPS}.

The violations of Bell inequalities provide interesting ways of quantifying inconsistence of quantum correlations with local classical description. Unfortunately, for a given quantum state the violation primarily depends on specially constructed Bell inequality. It turns out to be a daunting task to build Bell inequality for general quantum states except for special cases. Clauser-Horne-Shimony-Holt (CHSH) inequality as the first nontrivial example \cite{CHSH1} can be applied for testing the inconsistence of generic bipartite entangled pure states \cite{Gisin1,GP1,PR1} or special mixed states \cite{BMR} in the case of dichotomic inputs and outputs of each local observer. Similar result holds for multipartite entangled pure states \cite{LF,YCZ} using Hardy's inequality \cite{Hard} which provides a state-dependent proof of the quantum contextuality. However, Werner's nonseparable mixed states $\rho=v|\Phi\rangle\langle \Phi|+\frac{1-v}{4}\mathbbm{1}$ serves as a counterexample to Gisin's Theorem \cite{Gisin1} in the case of mixed states using CHSH inequality. As mixtures of entangled pure state $|\Phi\rangle$ with white noise, Werner states have been used to describe the state of each molecule consisting of active nuclear spins in NMR \cite{Brau} and may be useful for single-qubit teleportation \cite{Pop,Benn}. Werner states violate CHSH inequality for $v>\frac{1}{\sqrt{2}}$ while they are separable if and only if $v\leq \frac{1}{3}$. Moreover, Werner states admit a LHV model in the case of projective measurements for $v\leq \frac{1}{2}$ \cite{Werner} or general measurements for $v\leq \frac{5}{12}$ \cite{Bar}. The range of $v$ for which a LHV model can be constructed for Werner states has been further extended to $v\leq 0.66$ \cite{Pop1,Bar,AG}.  It is unknown whether Werner states admit a LHV model for $0.66\leq v \leq \frac{1}{\sqrt{2}}$, or they violate a special Bell inequality.

Inspired by this open problem, in this letter we investigate the undetectable problem of general Werner states. Although each entangled state can be separated from all separable states using a linear functional from Hahn-Banach theorem, however, Werner's example demonstrated that the separating mapping may be unavailable for some mixed entanglement in terms of LHV models. We focus on LHV model with dichotomic inputs and outputs of each local observer. For homogeneous linear Bell inequalities consisting of full correlations, we prove that Werner states of almost all generalized multipartite Greenberger-Horne-Zeilinger (GHZ) states cannot be completely detected. The main method is to compare the critical visibility in terms of Bell inequality and the critical parameter in terms of full separability. Different from the quantum bound of Bell inequality using singular value decomposition of coefficient matrix \cite{EKB}, we estimate general quantum and classical bounds for Werner states of all generalized multipartite GHZ states. Our result mainly relies on direct estimation methods using Bell operators derived from Bell inequalities, which are also different from operator space theory \cite{Ts,JPP}. Similar result holds for Werner states of entangled pure state. Some generalized GHZ states \cite{GHZ1} do not violate any Bell inequality involving two dichotomic observables per local party \cite{ZBL}, however, the result does not hold for sub-correlations \cite{CHSH1} which are considered by us. For mixed states, the only result of the undetectability holds for Werner state of the maximally entangled Einstein-Podolsky-Rosen (EPR) state \cite{ZB1,EPR1}. We present a general undetectability of homogeneous linear Bell inequalities consisting of dichotomic inputs and outputs. For a general linear Bell inequality, we provide an accessible method to test the undetectablity of Werner states of generalized multipartite GHZ states or arbitrary entangled pure state. Based on numeric evaluations, there exists nonzero measure of undetectable Werner states. These results are related to the old mathematical problem of estimating Grothendieck-type constants \cite{Ts,AG}, which are only useful for bipartite Bell inequalities.

\section*{Preliminaries}

In what follows, we focus on the scenario of dichotomic inputs and outputs of each local observer. A general $m$-qubit state is represented by density operator $\rho$ on Hilbert space $\mathbb{H}=\mathbb{H}^{\otimes m}_2$, where $\mathbb{H}_2$ denotes Hilbert space of single qubit states. $\rho$ is fully separable if there exist a normalized probability distribution $\{p_i, i=1, \cdots, n\}$ and density operators $\{\rho_{i}^j, i=1, \cdots, n; j=1, \cdots, m\}$ which are mixed states of the $i$-th subsystem, such that $\rho=\sum_{i=1}^np_i\otimes_{j=1}^m\rho_{i}^j$. Otherwise, $\rho$ is not fully separable. Without loss of generality, for each $j$ all states $\rho_{i}^j (j=1, \cdots, m)$ may be assumed to be rank-1 projections, i.e., ensembles of pure states. Mathematically, all fully separable states are convex polytopes of pure product states. The convexity implies the separability of fully separable states and each entangled state from the famous Hahn-Banach theorem.

Assume that the $j$-th space-like separated observer has two inputs labeled by $x_{j}$ (which equals to 0 or 1 for simplicity), where the corresponding outcomes are labeled by $a_{j}=\pm1$. The key of LHV model is the probability distribution of $a_{j}$ given $x_{j}$, that is, the conditional probability $P(a_{1},\cdots, a_{m}|x_{1}, \cdots, x_{m})$.

{\bf Definition 1}. A LHV model \cite{Bell1,Pop1} is described by a conditional probability
\begin{eqnarray}
P(a_{1},\cdots, a_{m}|x_{1}, \cdots, x_{m})=\int_{\Omega} d\mu(\lambda) \prod_{j=1}^mp_\lambda(a_{j}|x_{j},\lambda)
\label{eqn1}
\end{eqnarray}
for all dichotomic variables $a_{i}$s and $x_{i}$s, where $(\Omega, \mu(\lambda))$ is a probability space of hidden variable $\lambda$ with normalization condition $\int_\Omega d\mu(\lambda)=1$. $\{p_\lambda(a_{j}=\pm1|x_{j},\lambda)\}$ ($j=1, \cdots, m$) are $m$ probability distributions depending on hidden variable $\lambda$.

{\bf Definition 2}. A quantum model is described by a conditional probability
\begin{eqnarray}
P(a_{1},\cdots, a_{m}|x_1, \cdots, x_{m})={\rm Tr}(\otimes_{j=1}^m{\bf A}^{a_j}_{x_j}\rho)
\label{eqn2}
\end{eqnarray}
for all dichotomic variables $a_{i}$s and $x_{i}$s, where $\rho$ is a density operator on Hilbert space $\mathbb{H}^{\otimes n}_2$. For each $i$, $\{{\bf A}^{a_i}_{x_i}, a_i=\pm1, x_i=0,1\}$ consists of noncommuting Hermitian operators with eigenvalues no more than $1$, which represent quantum measurements on the $i$-th observer's system, i.e., $\|{\bf A}^{a_i}_{x_i}\|\leq 1$ for all $a_i$s and $x_i$s, $[{\bf A}^{a_i}_{x_i=0},{\bf A}^{a_i}_{x_i=1}]\not=0$ for each $a_i$, and $\sum_{a_i=\pm1}{\bf A}^{a_i}_{x_i}=\mathbbm{1}_2$ (identity operator on Hilbert space $\mathbb{H}_2$) for each $x_i$.

Quantum state $\rho$ used in Definition 2 is interpreted as shared quantum resource. When $\rho$ is fully separable, Quantum model defined in Definition 2 is a special case of LHV defined in Definition 1 with discrete distribution of $\lambda$, which may be obtained by measuring the subsystems with rank-1 projectors. Generally, a LHV model is represented by the following local hidden state \cite{Ts}:

{\bf Definition 3} A Local Hidden State (LHS) model is described by a conditional probability
\begin{eqnarray}
P(a_{1},\cdots, a_{m}|x_1, \cdots, x_{m})={\rm Tr}(\otimes_{i=1}^m{\bf A}^{a_i}_{x_i}\rho)
\label{eqn2}
\end{eqnarray}
for all dichotomic variables $a_{i}$s and $x_{i}$s, where $\rho$ is a density operator on Hilbert space $\mathbb{H}^{\otimes n}_2$, $\{{\bf A}^{a_i}_{x_i}\} (i=1, \cdots, m)$ consists of commuting measurement operators with eigenvalues no more than $1$, which represent projective (classical) measurements on the $i$-th observer's system, i.e., $\|{\bf A}^{a_i}_{x_i}\|\leq 1$ for all $a_i$s and $x_i$s, $[{\bf A}^{a_i}_{x_i=0},{\bf A}^{a_i}_{x_i=1}]=0$ for each $a_i$, and $\sum_{a_i=\pm1}{\bf A}^{a_i}_{x_i}=\mathbbm{1}_2$ for each $x_i$.

\section*{Results}

From Definition 1, define $\langle A_{x_{1}}\cdots A_{x_{m}} \rangle=\sum_{a_1, \cdots, a_m=-1, 1}(-1)^{\sum_{i=1}^m(a_{i}+1)/2}\frac{1}{2^m}P(a_1, \cdots, a_m|x_1, \cdots, x_m)$ as the expectation of outputs $a_i$s derived from $m$-partite correlations $P(a_1, \cdots, a_m|x_1, \cdots, x_m)$ with inputs $x_i$s, where we assume the uniform probability distribution for each input $x_i$. Let $\langle B\rangle=\sum_{x_{1},\cdots, x_{m}}\alpha_{x_1\cdots x_m}\langle A_{x_{1}}\cdots A_{x_{m}}\rangle$ be linear superposition of some expectations $\langle A_{x_{1}}\cdots A_{x_{m}}\rangle$. The corresponding Bell inequalities are given by $|\langle B\rangle|_c\leq c_1$ and $|\langle B \rangle|_q\leq c_2$, where $c_1$ denotes classical bound over all conditional probabilities in Eq.(1) in terms of LHV model in Definition 1 while $c_2$ denotes quantum bound over all conditional probabilities in Eq.(2) in terms of quantum model in Definition 2. The corresponding Bell operator is given by ${\cal B}=\sum_{x_{1},\cdots, x_{m}}\alpha_{x_1\cdots x_m} \sum_{a_1, \cdots, a_m=-1, 1}(-1)^{\sum_{i=1}^m(a_{i}+1)/2}{\bf A}^{a_1}_{x_{1}}\otimes\cdots \otimes {\bf A}^{a_m}_{x_{m}}$. From Definitions 2 and 3, we have $|\langle B\rangle|_{c}=\sup_{{\bf A}^{a_1}_{x_{1}}, \cdots, {\bf A}^{a_m}_{x_{m}},\rho}|{\rm Tr}({\cal B}\rho)|:=\|{\cal B}\|_{c}$ over all commuting Hermitian operators ${\bf A}^{a_i}_{x_{i}}$s and density operators $\rho$ on Hilbert space $\mathbb{H}_2^m$, while $|\langle B\rangle|_{q}=\sup_{{\bf A}^{a_1}_{x_{1}}, \cdots, {\bf A}^{a_m}_{x_{m}},\rho}|{\rm Tr}({\cal B}\rho)|:=\|{\cal B}\|_{q}$ over all general Hermitian operators ${\bf A}^{a_i}_{x_{i}}$s and density operators $\rho$ on Hilbert space $\mathbb{H}_2^m$.

Given an $m$-partite generalized GHZ state \cite{GHZ1} $|\Psi\rangle=\cos\theta|\vec{0}\rangle+\sin\theta|\vec{1}\rangle$ with $\theta\in (0,\frac{\pi}{2})$, the corresponding Werner state \cite{Werner} is defined as mixed state of $|\Psi\rangle\langle \Psi|$ and noise state $\frac{1}{2^m}\mathbbm{1}$ with the following form:
\begin{eqnarray}
\rho_v=\frac{1-v}{2^m}\mathbbm{1}+v|\Psi\rangle\langle \Psi|,
\label{eq1}
\end{eqnarray}
where $\mathbbm{1}$ denotes the identity operator on Hilbert space $\mathbb{H}^{\otimes m}_2$, $|\vec{0}\rangle=|0\rangle^{\otimes m}, |\vec{1}\rangle=|1\rangle^{\otimes m}$ and $v\in [0, 1]$. The weight $v$ may operationally represent the observed interferometric contrast in experiment \cite{Werner}. Our first result is a typical undetectable feature of Werner states. Formally, we prove the following theorem.

{\bf Theorem 1}. The following results hold:
\begin{itemize}
\item[(i)] For almost all generalized GHZ states, the corresponding Werner states defined in Eq.(\ref{eq1}) cannot be completely detected by homogeneous linear Bell inequalities;

\item[(ii)]For most of generalized GHZ states, the corresponding Werner states defined in Eq.(\ref{eq1}) cannot be completely detected by a general linear Bell inequality if the corresponding Bell operator ${\cal B}$ satisfies
\begin{eqnarray}
\|{\cal B}\|_c \geq \gamma_i \|{\cal B}_i\|_c
\label{eq3}
\end{eqnarray}
for $i=1, \cdots, m-1$, where all positive constants $\gamma_i$s satisfy
 \begin{eqnarray}
 \sum_{i=1}^{m-1}\frac{1}{\gamma_i}<\sqrt{12}(2^m-1).
 \label{eq2}
\end{eqnarray}
Here, ${\cal B}_i$ denotes partial Bell operator of ${\cal B}$ including the correlation of the $i$-th party but excluding all correlations of the $j$-th parties, $j=1, \cdots, i-1$;

\item[(iii)]For most of generalized GHZ states with $3\leq m\leq 6$, the corresponding Werner states defined in Eq.(\ref{eq1}) cannot be completely detected by general linear Bell inequalities.
\end{itemize}

Theorem 1 generalizes the undetectability of Werner state of the maximally entangled GHZ state \cite{Werner,ZB1}. In particular, the generic undetectability holds for homogeneous linear Bell inequalities. Here, {\it homogeneous} means that all terms of $m$-partite Bell inequality include full correlations of $m$ observers. {\it Almost} means for almost all $\theta\in (0, \frac{\pi}{2})$ (approximate full measure) Werner states defined in Eq.(\ref{eq1}) cannot be completely detected by any homogeneous linear Bell inequality, i.e., the critical visibility $v$ in terms of the violation of some linear Bell inequality is larger than the critical parameter $v$ in terms of the full separability. Hence, there are Werner states that are entangled but cannot be detected by Bell testing. The main idea is to estimate upper bounds of all homogeneous linear Bell inequalities (Appendices A, B and C). CHSH inequality \cite{CHSH1} and Mermin inequality \cite{Merm} are nontrivial examples that cannot completely detect almost all Werner states defined in Eq.(\ref{eq1}). A general linear Bell inequality may have sub-correlations involving less than $m$ observers and has no tight upper bound in terms of LHV model. The conditions presented in Eqs.(\ref{eq3}) and (\ref{eq2}) provide an accessible way to determine whether a given Werner state is undetectable or not. Some examples are shown in Appendix H.

Given an $m$-partite entangled pure state $|\Phi\rangle=\sum_{i_1, \cdots, i_m=0, 1}\alpha_{\vec{i}}|\vec{i}\rangle$ with $m\geq2$, Werner states \cite{Werner} of $|\Phi\rangle$ are defined by
\begin{eqnarray}
\rho_v=\frac{1-v}{2^m}\mathbbm{1}+v|\Phi\rangle\langle \Phi|,
\label{eq4}
\end{eqnarray}
where $\alpha_{\vec{i}}$ are complex coefficients satisfying $\sum_{\vec{i}}|\alpha_{\vec{i}}|^2=1$, $\vec{i}$ denote $m$-bit index vectors $i_1\cdots i_m$, and $v\in [0, 1]$. $\rho_v$ are density matrices for all $v\in [0, 1]$. Similar results hold for generalized Werner states.

{\bf Theorem 2}. The following results hold:
\begin{itemize}
\item[(i)] For almost all pure states, the corresponding Werner states defined in Eq.(\ref{eq4}) cannot be completely detected by homogeneous linear Bell inequalities;

\item[(ii)]There is a nonzero measure of entangled pure states whose Werner states defined in Eq.(\ref{eq4}) cannot be completely detected by a general linear Bell inequality, if the corresponding Bell operator ${\cal B}$ satisfies
 \begin{eqnarray}
\|{\cal B}\|_c \geq \gamma_i \|{\cal B}_i\|_c
\label{eq6}
\end{eqnarray}
for $i=1, \cdots, m-1$, where positive constants $\gamma_i$ satisfy
\begin{eqnarray}
\sum_{i=1}^{m-1}\frac{1}{\gamma_i}<\frac{1}{\sqrt{3}}{\rm Poly}(m)-1.
\label{eq5}
\end{eqnarray}
Here, ${\cal B}_i$ denotes partial Bell operator including the correlation of the $i$-th party but excluding all correlations of the $j$-th parties ($j=1, \cdots, i-1$), and ${\rm Poly}(\cdot)$ denotes any polynomial function;

\item[(iii)]There is a nonzero measure of $m$-partite pure states with $3\leq m\leq 6$ with the property that the corresponding Werner states defined in Eq.(\ref{eq4}) cannot be completely detected by general linear Bell inequalities.

\end{itemize}

Theorem 2 generalizes the undetectability of Werner state of bipartite states \cite{Pop1,Bar,AG} and special multipartite states \cite{Werner,ZB1,BFFH}. Similar to Theorem 1, there are many instances of undetectable Werner states of general pure states in the case of homogeneous linear Bell inequalities such as CHSH inequality \cite{CHSH1} and Mermin inequality \cite{Merm}. Different from generalized GHZ states, there is no tight upper bound of the critical parameter $v$ for each pure state $|\Phi\rangle$ in terms of the full separability (Appendices D, F and G). We cannot prove the typical undetectability for Werner states of general pure states in the case of general linear Bell inequalities. Although Grothendieck-type constants may be used for bipartite, it is unclear how to extend the result to multipartite states \cite{Ts,AG}. Our conditions given in Eqs.(\ref{eq6}) and (\ref{eq5}) may be useful for determining whether a general Werner state is detectable or not for given linear Bell inequalities.

\section*{Discussion}

Werner states as special mixed states can be geometrically described as a curve of states in density operator space, which is from a fully separable mixed state to an entangled pure state \cite{Werner}. There always exists a critical point separating fully separable states and entangled states on the curve. These critical points generally have different mixing parameters $v$ that depend on the given pure states. The relative critical visibility as the ratio of quantum bound and classical bound of Bell inequalities characterizes the delectability of these inequalities. Although all pure states admit a universal violation \cite{YCZ}, homogeneous Bell inequalities can only provide bounded violations that are incompatible with infinitely small critical parameter $v$ in terms of the full separability. We conjecture that this incompatibility holds for general linear Bell inequalities even though there are unbounded violations \cite{JPP}. The possible reason is that almost all critical points require exponential violations \cite{Deng1} (Appendices E, F) going beyond polynomial violations \cite{JPP}. Unfortunately, our algorithmic proof (Appendix F) can only provide evidences for small $m$ due to high computational complexity. It is an open problem to find new methods for estimating unbounded violations. We further conjecture similar results for multipartite high-dimensional states.

In summary, we have shown that Gisin's theorem \cite{Gisin1} does not hold for most Werner states in the case of homogeneous Bell inequalities consisting of full correlations, where each observer can choose two dichotomic observables. For a general linear Bell inequality without tight classical bound, we presented accessible conditions to determine the detectability in the case of general Werner states. Our conditions involve only classical bounds of $m-1$ restricted Bell operators. Moreover, with these conditions we numerically proved the undetectable Werner states for small $m$. Of course, the presented incompatibility may be resolved for different assumptions such as multiple observables per local observer \cite{KC}, which are also interesting problems. Our results may shed a new light on entanglement testing, quantum supremacy or quantum communication \cite{SG}.

\section*{Acknowledgements}

We thank the helpful discussions of Luming Duan, Yaoyun Shi, Yuan Su, Xiubo Chen and Yixian Yang. Authors thank for interesting suggestions by Joseph Bowles. This work was supported by the National Natural Science Foundation of China (Nos.61772437,61702427), Sichuan Youth Science and Technique Foundation (No.2017JQ0048), Fundamental Research Funds for the Central Universities (No.2682014CX095), Chuying Fellowship, and EU ICT COST CryptoAction (No.IC1306).

\section*{Appendix A: Ideal bound of the critical visibility in the case of linear Bell inequality}

We firstly introduce some notation. Let $\vec{i}$ denote an $m$-bit index vector $i_1\cdots i_m$, i.e, the concatenation of $m$ bits $i_1, \cdots, i_m$. Denote $\vec{i}_c=\vec{1}\oplus \vec{i}$ as binary complement of $\vec{i}$, where $\vec{1}$ represents $m$-bit of $1\cdots 1$ and addition is mod 2 per bit.

Given an $m$-partite generalized GHZ state \cite{GHZ} $|\Psi\rangle=\cos\theta|\vec{0}\,\rangle+\sin\theta|\vec{1}\,\rangle$ with $\theta\in (0,\frac{\pi}{2})$, Werner state \cite{Wer} is a mixed state of pure state $|\Psi\rangle\langle \Psi|$ and noise state (maximally mixed state) $\frac{1}{2^m}$ with the following form:
\begin{align*}
\rho_v=\frac{1-v}{2^m}\mathbbm{1}+v|\Psi\rangle\langle \Psi|,
\tag{A1}
\end{align*}
where $\mathbbm{1}$ denotes the identity operator on Hilbert space $\mathbb{H}^{\otimes m}_2$, $\mathbb{H}_2$ denotes Hilbert space of single qubit states, and $v\in [0, 1]$. It easily proves that $\rho_v$ are density matrices (positive semidefinite) for all $v\in [0,1]$.

Define $\langle A_{x_{1}}\cdots A_{x_{m}} \rangle=\sum_{a_1, \cdots, a_m=-1, 1}(-1)^{\sum_{i=1}^m(a_{i}+1)/2}P(a_1, \cdots, a_m|x_1, \cdots, x_m)$ as the expectation of $m$-partite measurements $A_{x_{1}}\cdots A_{x_{m}}$ with inputs $x_1, \cdots, x_m$, where we assume the uniform probability distribution for each input $x_i$. Denote $\langle B\rangle=\sum_{x_{1},\cdots, x_{m}}\alpha_{x_1\cdots x_m} \langle A_{x_{1}}\cdots A_{x_{m}}\rangle$. The corresponding Bell inequalities are given by $|\langle B\rangle|_c\leq c_1$ and $|\langle B \rangle|_q\leq c_2$, where $c_1$ denotes classical bound over all conditional probabilities shown in Eq.(1) in terms of LHV model in Definition 1 while $c_2$ denotes quantum bound over all conditional probabilities given in Eq.(2) in terms of quantum model in Definition 2. The corresponding Bell operator is given by ${\cal B}=\sum_{x_{1},\cdots, x_{m}}\alpha_{x_1\cdots x_m} \sum_{a_1, \cdots, a_m=-1, 1}(-1)^{\sum_{i=1}^m(a_{i}+1)/2}{\bf A}^{a_1}_{x_{1}}\otimes\cdots \otimes {\bf A}^{a_m}_{x_{m}}$. From Definitions 2 and 3, we have $|\langle B\rangle|_{c}=\sup_{{\bf A}^{a_1}_{x_{1}}, \cdots, {\bf A}^{a_m}_{x_{m}},\rho}|{\rm Tr}({\cal B}\rho)|:=\|{\cal B}\|_{c}$ over all commuting Hermitian operators ${\bf A}^{a_i}_{x_{i}}$s and density operator $\rho$ on Hilbert space $\mathbb{H}_2^m$, while $|\langle B\rangle|_{q}=\sup_{{\bf A}^{a_1}_{x_{1}}, \cdots, {\bf A}^{a_m}_{x_{m}},\rho}|{\rm Tr}({\cal B}\rho)|:=\|{\cal B}\|_{q}$ over all possible Hermitian operators ${\bf A}^{a_i}_{x_{i}}$s and density operator $\rho$ on Hilbert space $\mathbb{H}_2^m$.

The proof of Theorem 1 is completed by the following four sections. This section contributes the lower bound of $v$ for which $\rho_v$ violates a linear Bell inequality.

Given a Bell operator ${\cal B}$ derived from a linear $m$-partite Bell inequality, assume that
\begin{align}
\| {\cal B}\|_c\leq c_1
\tag{A2}
\end{align}
in terms of LHS model, and
\begin{align*}
\|{\cal B}\|_q\leq c_2
\tag{A3}
\end{align*}
in terms of quantum model, where $\vec{\bf A}^{c}$ and $\vec{\bf A}^{q}$ denote $({\bf A}^{c}_{1,i_1}, \cdots, {\bf A}^{c}_{1,i_m})$ and $({\bf A}^{q}_{1,i_1}, \cdots, {\bf A}^{q}_{1,i_m})$ respectively, ${\bf A}^{c}_{j,i_j}$ denote commuting Hermitian operators of the $j$-th party, and ${\bf A}^{q}_{j,i_j}$ denote general Hermitian operators of the $j$-th party. To detect quantum entanglement $\rho_v$, $c_i$s satisfy $c_2>c_1$.

{\bf Lemma 1}. The critical visibility $v^*$ for which $\rho_v$ defined in Eq.(A1) violates linear Bell inequality shown in Eq.(A2) is given by
\begin{align*}
v^*\geq \frac{2^mc_1-c_1}{2^mc_2-c_1}.
\tag{A4}
\end{align*}

{\bf Proof}. Assume that Werner state $\rho_v$ defined in Eq.(A1) violates linear Bell inequality shown in Eq.(A2). There exist general Hermitian measurements $\vec{\bf A}^q$ and density operator $\rho$ on Hilbert space $\mathbb{H}_2^m$ such that \begin{align*}
{\rm Tr} ({\cal B}_{\vec{\bf A}^q}\rho)=&\frac{1-v}{2^m}{\rm Tr}({\cal B}_{\vec{\bf A}^q} \mathbbm{1})+v{\rm Tr}({\cal B}_{\vec{\bf A}^q} |\Psi\rangle\langle \Psi|)
\\
>& c_1,
\tag{A5}
\end{align*}
where we assume ${\rm Tr} ({\cal B}_{\vec{\bf A}^q}\rho)>0$; Otherwise, we replace ${\cal B}_{\vec{\bf A}^q}$ with $-{\cal B}_{\vec{\bf A}^q}$, where ${\cal B}_{\vec{\bf A}^q}$ is the Bell operator derived from general Hermitian measurement operators $\vec{\bf A}^q$. Since ${\rm Tr}({\cal B}_{\vec{\bf A}^q} \mathbbm{1})\leq c_1$ ($\frac{1}{2^m}\mathbbm{1}$ is the maximally mixed state which is fully separable) and ${\rm Tr}({\cal B}_{\vec{\bf A}^q}|\Psi\rangle\langle \Psi|)\leq c_2$ from Eqs.(A2) and (A3), Eq.(A5) implies that
\begin{align*}
v> &\frac{2^mc_1-{\rm Tr} ({\cal B}_{\vec{\bf A}^q}\mathbbm{1})}{2^m{\rm Tr} ({\cal B}_{\vec{\bf A}^q}|\Psi\rangle\langle \Psi|)-{\rm Tr} ({\cal B}_{\vec{\bf A}^q}\mathbbm{1})}
\\
\geq&\frac{2^mc_1-{\rm Tr} ({\cal B}_{\vec{\bf A}^q}\mathbbm{1})}{2^mc_2-{\rm Tr} ({\cal B}_{\vec{\bf A}^q}\mathbbm{1})}
\\
\geq&\frac{2^mc_1-c_1}{2^mc_2-c_1},
\tag{A6}
\end{align*}
where Eq.(A6) is from the inequality $\frac{a-x}{b-x}\geq \frac{a-x'}{b-x'}$ for $x,x'$ satisfying $0\leq x\leq x'\leq \min\{a, b\}$. $\Box$

\section*{Appendix B: Relative critical visibility in the case of homogeneous linear Bell inequalities}

This section contributes the relative critical visibility, i.e., the ratio of the quantum upper bound and the classical upper bound of homogeneous linear Bell inequalities. In what follows, we focus on the case of dichotomic inputs and outputs for each observer. Two measurement operators ${\bf A}_{i,x_i=0}$ and ${\bf A}_{i,x_i=1}$ are used for the $i$-th observer, $i=1, \cdots, m$. Assume that each measurement operator ${\bf A}_{i,x_i}$ has $\pm1$ eigenvalues. $\langle{\bf A}_{i, x_i}\rangle$ denotes the expectation of outputs of the $i$-th observer. With these denotations, from Definitions 2 and 3 a general homogeneous linear Bell inequality may be given by
\begin{align*}
\sup_{\vec{\bf A}^{c(q)},\rho}|{\rm Tr} (\sum_{i_1,\cdots, i_m=0,1}\alpha_{\vec{i}}{\bf A}^{c(q)}_{1, i_1}\otimes \cdots \otimes {\bf A}^{c(q)}_{m, i_m}\rho)|
\leq &c_1(c_2),
\tag{B1}
\end{align*}
where supremum is over all measurement operators $\vec{\bf A}^{c(q)}$ and density operator $\rho$ on Hilbert space $\mathbb{H}_2^m$ \cite{Tsi1,Tsi2}, $c_1$ and $c_2$ denote the respective upper bound in the case of commuting Hermitian operators ${\bf A}^{c}_{i,j}$s and general Hermitian operators ${\bf A}^q_{i,j}$s, and $\alpha_{\vec{i}}$ are real coefficients. Here, general Hermitian operators include quantum observeables and commuting Hermitian operators. The corresponding Bell operator is given by
\begin{align*}
{\cal B}=\sum_{i_1,\cdots, i_m}\alpha_{\vec{i}}{\bf A}_{1, i_1}\otimes \cdots \otimes {\bf A}_{m, i_m},
\tag{B2}
\end{align*}
which is a homogeneous linear Hermitian operator. In this paper, {\it homogeneous} means that all joint measurement operators ${\bf A}^{c(q)}_{1, i_1}\otimes \cdots \otimes {\bf A}^{c(q)}_{m, i_m}$ of ${\cal B}$ include exactly $m$ measurement operators ${\bf A}^{c(q)}_{1, i_1}, \cdots, {\bf A}^{c(q)}_{m, i_m}$, or equivalently, the corresponding Bell inequality includes all full correlations.

Our goal in this section is to estimate the relative critical visibility $r_{cq}$ by taking use of  Hermitian operator ${\cal B}$ as
\begin{align*}
r_{cq}=\frac{\|{\cal B}\|_{q}}{\|{\cal B}\|_{c}},
\tag{B3}
\end{align*}
where $\|{\cal B}\|_{c}:=\sup_{\vec{\bf A}^{c},\rho}|{\rm Tr} (\sum_{i_1,\cdots, i_m=0,1}\alpha_{\vec{i}}{\bf A}^{c}_{1, i_1}\otimes \cdots \otimes {\bf A}^{c}_{m, i_m}\rho)|$ denotes the norm of Bell operator ${\cal B}$ over commuting Hermitian operators ${\bf A}^{c}_{j, i_j}$s and density operators $\rho$ while $\|{\cal B}\|_{q}:=\sup_{\vec{\bf A}^{q},\rho}|{\rm Tr} (\sum_{i_1,\cdots, i_m=0,1}\alpha_{\vec{i}}{\bf A}^{q}_{1, i_1}\otimes \cdots \otimes {\bf A}^{q}_{m, i_m}\rho)|$ denotes the norm of Bell operator ${\cal B}$ over general Hermitian operators ${\bf A}^{q}_{j, i_j}$s and density operators $\rho$. Here, $\|{\cal B}\|_{c}\not=0$ for any nonzero Bell operators. The upper bound of $r_{cq}$ for homogeneous Bell operator ${\cal B}$ is shown in Lemma 2. Typical examples are CHSH inequality \cite{CHSH} and Mermin inequality \cite{Mermin}.

{\bf Lemma 2}. For all homogeneous Bell operators of $m$-multipartite, the relative critical visibility $r_{cq}$ defined in Eq.(B3) satisfies
\begin{align*}
r_{cq}\leq \sqrt{3},
\tag{B4}
\end{align*}
or $r_{cq}\leq \sqrt{\frac{5}{2}}$ when quantum observables are restricted to anticommuting observables.

{\bf Proof}. We first consider $m=2$ to explain the main idea. And then, we extend the proof for $m>2$.

Case 1. $m=2$.

Let ${\bf A}_{1,0}$ and ${\bf A}_{1,1}$ be dichotomic measurement operators of Alice, and ${\bf A}_{2,0}$ and ${\bf A}_{2,1}$ be dichotomic measurement operators of Bob. ${\bf A}_{i,j}$ are Hermitian operators with eigenvalues $\lambda$ satisfying $|\lambda|\leq 1$, i.e., $\|{\bf A}_{i,j}\|\leq 1$ ($\|\cdot\|$ denotes the norm of operators in terms of quantum model). For simplicity, we assume ${\bf A}_{i,j}^2=\mathbbm{1}_2$ (the identity operator on single qubit state), i.e, the outcomes of observables are associated to projective measurements \cite{Tsi1,Tsi2}.

Given a bipartite homogeneous linear Bell operator ${\cal B}$, we have
\begin{align*}
{\cal B}=&\sum_{i,j=0}^1\alpha_{ij} {\bf A}_{1,i}\otimes{\bf A}_{2,j}
\\
=&\sum_{i=1}^4\alpha_{i} {\bf A}_{i}\otimes{\bf B}_{i},
\tag{B5}
\end{align*}
where ${\bf A}_{1}={\bf A}_2={\bf A}_{1,0}$ and ${\bf A}_3={\bf A}_4={\bf A}_{1,1}$, ${\bf B}_1={\bf B}_3=\frac{1}{2}({\bf A}_{2,0}+{\bf A}_{2,1})$ and ${\bf B}_2={\bf B}_4=\frac{1}{2}({\bf A}_{2,0}-{\bf A}_{2,1})$, $\alpha_1=\alpha_{00}+\alpha_{01}$, $\alpha_2=\alpha_{00}-\alpha_{01}$, $\alpha_3=\alpha_{10}+\alpha_{11}$ and $\alpha_4=\alpha_{10}-\alpha_{11}$. From Eq.(B5) we obtain
\begin{align*}
{\cal B}^2=&\sum_{i,j=1}^4\alpha_{i} \alpha_{j}({\bf A}_i{\bf A}_j)\otimes ({\bf B}_i{\bf B}_j)
\\
=&(\alpha_1{\bf A}_1+\alpha_3{\bf A}_3)^2\otimes {\bf B}_1^2
+(\alpha_2{\bf A}_2+\alpha_4{\bf A}_4)^2\otimes {\bf B}_2^2
\\
&+\frac{1}{4}(\alpha_1{\bf A}_1+\alpha_3{\bf A}_3)(\alpha_2{\bf A}_2+\alpha_4{\bf A}_4)\otimes [{\bf A}_{2,1},{\bf A}_{2,0}]^2
\\
&+\frac{1}{4}(\alpha_2{\bf A}_2+\alpha_4{\bf A}_4)(\alpha_1{\bf A}_1+\alpha_3{\bf A}_3)\otimes [{\bf A}_{2,0},{\bf A}_{2,1}]^2,
\tag{B6}
\end{align*}
where $[{\bf X}, {\bf Y}]={\bf X}{\bf Y}-{\bf Y}{\bf X}$.

In what follows, assume that $\alpha_1\geq0$ and $\alpha_2\geq0$; Otherwise, by replacing $\alpha_i$ with $-\alpha_i$, the value of the right side of Eq.(B6) is invariable. For commuting Hermitian operators ${\bf A}^c_{i,j}$, ${\bf A}_i$ and ${\bf B}_i$ are commuting. Given a density operator $\rho$ on Hilbert space $\mathbb{H}_2^2$, Eq.(B6) implies that
\begin{align*}
{\rm Tr}({\cal B}^2\rho)
={\rm Tr}((\alpha_1{\bf A}_1+\alpha_3{\bf A}_3)^2\otimes{\bf B}_1^2\rho) +{\rm Tr}((\alpha_2{\bf A}_2+\alpha_4{\bf A}_4)^2\otimes{\bf B}_2^2\rho),
\tag{B7}
\end{align*}
where Eq.(B7) is from the equalities $[{\bf B}_i,{\bf B}_j]=0$ with $i=1, 3$ and $j=2, 4$.  Two different cases will be discussed as follows:
\begin{itemize}
\item[(i)] $\alpha_3\alpha_4\geq0$. Eq.(B7) implies
\begin{align*}
|{\rm Tr}({\cal B}^2\rho)|\leq&(\alpha_1+|\alpha_3|)^2\beta_1^2
+(\alpha_2+|\alpha_4|)^2\beta_2^2,
\tag{B8}
\end{align*}
where Eq.(B8) is from the inequalities $\|{\bf A}_i\|\leq 1$ ($i=1, \cdots, 4$), $\beta^2_{1}={\rm Tr}({\bf B}_1^2\rho_B)=\frac{1}{2}+\frac{1}{4}{\rm Tr}({\bf A}^c_{2,0}{\bf A}^c_{2,1}\rho_B+{\bf A}^c_{2,1}{\bf A}^c_{2,0}\rho_B)$ and $\beta_{2}^2={\rm Tr}({\bf B}_2^2\rho_B)=\frac{1}{2}-\frac{1}{4}{\rm Tr}({\bf A}^c_{2,0}{\bf A}^c_{2,1}\rho_B+{\bf A}^c_{2,1}{\bf A}^c_{2,0}\rho_B)$ with the reduced state $\rho_{B}={\rm Tr}_A(\rho)$ of Bob. The equality of Eq.(B8) holds for separable density matrices $\rho=\rho_A\otimes \rho_B$, and identity measurement operators ${\bf A}_1={\bf A}_2=\mathbbm{1}_2$ and ${\bf A}_i=\frac{\alpha_i}{|\alpha_i|}\mathbbm{1}_2$ ($i=3, 4$) if $\alpha_i\not=0$. Now, two subcases will be given as follows:
\begin{itemize}
\item[(a)] If $\alpha_1+|\alpha_3|\geq \alpha_2+|\alpha_4|$, Eq.(B8) implies that
\begin{align*}
\sup_{\vec{\bf A}^c, \rho}|{\rm Tr}({\cal B}^2\rho)|^{\frac{1}{2}}=\alpha_1+|\alpha_3|,
\tag{B9}
\end{align*}
where the supremum is achieved when $\rho=\frac{1}{4}\mathbbm{1}$ and ${\bf A}^c_{2,0}={\bf A}^c_{2,1}=\mathbbm{1}_2$. From Schwartz inequality $|{\rm Tr}({\cal B}\rho)| \leq \sqrt{|{\rm Tr}({\cal B}^2\rho)|}$ we have
\begin{align*}
\sup_{\vec{\bf A}^c,\rho} |{\rm Tr}({\cal B}\rho)| &\leq \sup_{\vec{\bf A}^c,\rho}\sqrt{|{\rm Tr}({\cal B}^2\rho)|}
\\
&=\alpha_1+|\alpha_3|,
\tag{B10}
\end{align*}
where the equality holds for $\rho=\frac{1}{4}\mathbbm{1}$, ${\bf A}_1={\bf A}_2={\bf A}^c_{2,0}={\bf A}^c_{2,1}=\mathbbm{1}_2$, ${\bf A}_3=\frac{\alpha_3}{|\alpha_3|}\mathbbm{1}_2$ with $\alpha_3\not=0$, and ${\bf A}_4=\frac{\alpha_4}{|\alpha_4|}\mathbbm{1}_2$ with $\alpha_4\not=0$.

\item[(b)] If $\alpha_1+|\alpha_3|\leq \alpha_2+|\alpha_4|$, Eq.(B8) implies that
\begin{align*}
\sup_{\vec{\bf A}^c,\rho}|{\rm Tr}({\cal B}\rho)|=\alpha_2+|\alpha_4|,
\tag{B11}
\end{align*}
where the supremum is achieved when $\rho=\frac{1}{4}\mathbbm{1}$ and ${\bf A}^c_{2,0}=-{\bf A}^c_{2,1}=\mathbbm{1}_2$. From Schwartz inequality we have
\begin{align*}
\sup_{\vec{\bf A}^c,\rho}|{\rm Tr}({\cal B}\rho)| &\leq \sup_{\vec{\bf A}^c,\rho}\sqrt{|{\rm Tr}({\cal B}^2\rho)|}
\\
&=\alpha_2+|\alpha_4|,
\tag{B12}
\end{align*}
where the equality holds for $\rho=\frac{1}{4}\mathbbm{1}$, ${\bf A}_1={\bf A}_2={\bf A}^c_{2,0}=-{\bf A}^c_{2,1}=\mathbbm{1}_2$, ${\bf A}_3=\frac{\alpha_3}{|\alpha_3|}\mathbbm{1}_2$ with $\alpha_3\not=0$, and ${\bf A}_4=\frac{\alpha_4}{|\alpha_4|}\mathbbm{1}_2$ with $\alpha_4\not=0$.

\end{itemize}

From Eqs.(B10) and (B12), the norm of Bell operator ${\cal B}$ in the case of commuting Hermitian measurements is given by
\begin{align*}
\|{\cal B}\|_c=&\sup_{\vec{\bf A}^c,\rho}|{\rm Tr}({\cal B}\rho)|
\\
=&\max\{\alpha_1+|\alpha_3|, \alpha_2+|\alpha_4|\}.
\tag{B13}
\end{align*}

\item[(ii)] $\alpha_3\alpha_4\leq0$. Eq.(B7) implies that
\begin{align*}
|{\rm Tr}( {\cal B}^2\rho)|\leq&\max\{(\alpha_1+|\alpha_3|)^2\beta_1^2
+(\alpha_2-|\alpha_4|)^2\beta_2^2,(\alpha_1-|\alpha_3|)^2\beta_1^2
+(\alpha_2+|\alpha_4|)^2\beta_2^2\},
\tag{B14}
\\
\leq & \max\{(\alpha_1+|\alpha_3|)^2, (\alpha_2-|\alpha_4|)^2, (\alpha_1-|\alpha_3|)^2, (\alpha_2+|\alpha_4|)^2\}
\tag{B15}
\\
=&  \max\{(\alpha_1+|\alpha_3|)^2, (\alpha_2+|\alpha_4|)^2\},
\tag{B16}
\end{align*}
where Eq.(B14) is from the inequalities $\|{\bf A}_i\|\leq 1$ ($i=1, \cdots, 4$), Eq.(B15) is from the inequalities $(\alpha_1+|\alpha_3|)^2\beta_1^2
+(\alpha_2-|\alpha_4|)^2\beta_2^2\leq \max\{(\alpha_1+|\alpha_3|)^2, (\alpha_2-|\alpha_4|)^2\}$ and $(\alpha_1-|\alpha_3|)^2\beta_1^2
+(\alpha_2+|\alpha_4|)^2\beta_2^2\leq \max\{(\alpha_1-|\alpha_3|)^2, (\alpha_2-|\alpha_4|)^2\}$ which may be proved similarly as those for Eqs.(B9)-(B12), and $\beta_{j}$ are defined in Eq.(B8). The equality of Eq.(B14) holds for ${\bf A}_1={\bf A}_2=\mathbbm{1}_2$, ${\bf A}_3=\frac{\alpha_3}{|\alpha_3|}\mathbbm{1}_2$ ($\alpha_3\not=0$) and ${\bf A}_4=-\frac{\alpha_4}{|\alpha_4|}\mathbbm{1}_2$ ($\alpha_4\not=0$), or ${\bf A}_1={\bf A}_2=\mathbbm{1}_2$, ${\bf A}_3=-\frac{\alpha_3}{|\alpha_3|}\mathbbm{1}_2$ and ${\bf A}_4=\frac{\alpha_4}{|\alpha_4|}\mathbbm{1}_2$. The equality of Eq.(B15) holds for the following cases respectively:
\begin{itemize}
\item[(a)] ${\bf A}_{2,0}={\bf A}_{2,1}=\mathbbm{1}_2$ and $(\alpha_1+|\alpha_3|)^2\geq (\alpha_2-|\alpha_4|)^2$;

\item[(b)] ${\bf A}_{2,0}=-{\bf A}_{2,1}=\mathbbm{1}_2$ and $(\alpha_1+|\alpha_3|)^2\leq (\alpha_2-|\alpha_4|)^2$;

\item[(c)] ${\bf A}_{2,0}={\bf A}_{2,1}=\mathbbm{1}_2$ and $(\alpha_1-|\alpha_3|)^2\geq (\alpha_2+|\alpha_4|)^2$;

\item[(d)] ${\bf A}_{2,0}=-{\bf A}_{2,1}=\mathbbm{1}_2$ and $(\alpha_1-|\alpha_3|)^2\leq (\alpha_2+|\alpha_4|)^2$.
\end{itemize}

Similar to Eqs.(B10) and (B12), we have the same equality as that given in Eq.(B13).
\end{itemize}

Now we estimate the norm of Bell operator ${\cal B}$ defined in Eq.(B5) in the case of quantum model. For general Hermitian operators ${\bf A}_{i,j}$ and density operator $\rho$, from Eq.(B6) and Schwartz inequality, we have
\begin{align*}
|{\rm Tr}({\cal B}\rho)|\leq &|{\rm Tr}({\cal B}^2\rho)|^{\frac{1}{2}}
\\
\leq &|{\rm Tr}((\alpha_1{\bf A}_1+\alpha_3{\bf A}_3)^2\otimes {\bf B}_1^2\rho)
+{\rm Tr}((\alpha_2{\bf A}_2+\alpha_4{\bf A}_4)^2\otimes{\bf B}_2^2\rho)
\\
&+{\rm Tr}((\alpha_1{\bf A}_1+\alpha_3{\bf A}_3)(\alpha_2{\bf A}_2+\alpha_4{\bf A}_4)\rho_A)
\\
&+{\rm Tr}((\alpha_2{\bf A}_2+\alpha_4{\bf A}_4)(\alpha_1{\bf A}_1+\alpha_3{\bf A}_3)\rho_A)|^{\frac{1}{2}}
\tag{B17}
\\
\leq &\sqrt{(|\alpha_1|+|\alpha_3|)^2\beta_1^2
+(|\alpha_2|+|\alpha_4|)^2\beta_2^2
+2(|\alpha_1|+|\alpha_3|)(|\alpha_2|+|\alpha_4|)}
\tag{B18}
\\
\leq &\max\{\sqrt{(|\alpha_1|+|\alpha_3|)^2
+2(|\alpha_1|+|\alpha_3|)(|\alpha_2|+|\alpha_4|)},
\\
&\qquad \sqrt{(|\alpha_2|+|\alpha_4|)^2
+2(|\alpha_1|+|\alpha_3|)(|\alpha_2|+|\alpha_4|)}\},
\tag{B19}
\end{align*}
where Eq.(B17) is from the inequalities $\|[{\bf B}_i, {\bf B}_j]\|\leq 2$ with $i=1, 3$ and $j=2, 4$, Eq.(B18) is from the inequalities $\|{\bf A}_i\|\leq 1$ ($i=1, \cdots, 4$); Eq.(B19) is from the fact that $\beta_i^2$ defined in Eq(B8) have invariable upper bounds for commuting or general Hermitian operators ${\bf B}_i$. Here, $\rho_A$ denotes the reduced state of Alice. So, we have
\begin{align*}
\|{\cal B}\|_q=&\sup_{\vec{\bf A}^q,\rho}|{\rm Tr}({\cal B}\rho)|
\\
\leq&\max\{\sqrt{(\alpha_1+|\alpha_3|)^2
+2(\alpha_1+|\alpha_3|)(\alpha_2+|\alpha_4|)},
\\
&\qquad \sqrt{(\alpha_2+|\alpha_4|)^2
+2(\alpha_1+|\alpha_3|)(\alpha_2+|\alpha_4|)}\}
\\
\leq &\max\{ \sqrt{3}(\alpha_1+|\alpha_3|), \sqrt{3}(\alpha_2+|\alpha_4|)\},
\tag{B20}
\end{align*}
where we have taken use of $2(\alpha_1+|\alpha_3|)(\alpha_2+|\alpha_4|)\leq (\alpha_1+|\alpha_3|)^2+(\alpha_2+|\alpha_4|)^2$.

Eqs.(B3), (B13) and (B20) imply that the upper bound of $r_{cq}$ defined in Eq.(B3) is given by
\begin{align*}
r_{cq}
\leq&\frac{\max\{ \sqrt{3}(\alpha_1+|\alpha_3|), \sqrt{3}(\alpha_2+|\alpha_4|)\}}{\max\{(\alpha_1+|\alpha_3|), (\alpha_2+|\alpha_4|)\}}
\\
=& \sqrt{3}.
\tag{B21}
\end{align*}

Especially, for anticommuting operators (Pauli matrices) ${\bf A}_{i,j}$, we have $\beta_1^2=\beta_2^2=\frac{1}{2}$. Eq.(B18) implies that
\begin{align*}
\sup_{\vec{\bf A}^q,\rho}|{\rm Tr}({\cal B}\rho)|
\leq &\sqrt{\frac{1}{2}(\alpha_1+|\alpha_3|)^2
+\frac{1}{2}(\alpha_2+|\alpha_4|)^2
+2(\alpha_1+|\alpha_3|)(\alpha_2+|\alpha_4|)}
\\
\leq &\max\{\sqrt{\frac{5}{2}}(\alpha_1+|\alpha_3|),
\sqrt{\frac{5}{2}}(\alpha_2+|\alpha_4|)\}.
\tag{B22}
\end{align*}
From Eqs.(B3), (B13) and (B22), we get an upper bound of the relative critical visibility as
\begin{align*}
r_{cq}
\leq&\frac{\max\{\sqrt{\frac{5}{2}}(\alpha_1+|\alpha_3|), \sqrt{\frac{5}{2}}(\alpha_2+|\alpha_4|)\}}{\max\{(\alpha_1+|\alpha_3|), (\alpha_2+|\alpha_4|)\}}
\\
=&\sqrt{\frac{5}{2}}.
\tag{B23}
\end{align*}

Case 2. $m>2$.

Consider an $m$-partite homogeneous linear Bell operator as
\begin{align*}
{\cal B}_m=&\sum_{\vec{i}}\alpha_{\vec{i}} {\bf A}_{1,i_1}\otimes \cdots \otimes {\bf A}_{m,i_m}
\\
:=&\sum_{i=1}^{2^m}\alpha_{i} \tilde{\bf A}_{i} \otimes {\bf B}_{i}
\tag{B24}
\end{align*}
where $\tilde{\bf A}_i:={\bf A}_{1,j_1} \otimes \cdots \otimes{\bf A}_{m-1,j_{m-1}}$ in which $j_1\cdots j_{m-1}$ are the first $m-1$ bits (from left to right) of the binary representation of integer $i$, $\alpha_{i}=\alpha_{j_1\cdots j_{m-1}0}+\alpha_{j_1\cdots j_{m-1}1}$ for odd integer $i$ with binary representation $j_1\cdots j_{m-1}0$, $\alpha_{i}=\alpha_{j_1\cdots j_{m-1}0}-\alpha_{j_1\cdots j_{m-1}1}$ for even integer $i$ with binary representation $j_1\cdots j_{m-1}1$, ${\bf B}_j=\frac{1}{2}({\bf A}_{m,0}+{\bf A}_{m,1})$ for all odd integers $j\leq 2^m$, ${\bf B}_j=\frac{1}{2}({\bf A}_{m,0}-{\bf A}_{m,1})$ for all even integers $j\leq 2^m$. From Eq.(B24) we have
\begin{align*}
{\cal B}_m^2=&\sum_{i,j=1}^{2^m}\alpha_{i} \alpha_{j} (\tilde{\bf A}_i\tilde{\bf A}_j)\otimes ({\bf B}_i{\bf B}_j)
\\
=&(\sum_{{\rm odd\,} i}\alpha_i\tilde{\bf A}_i)^2\otimes {\bf B}_1^2
+(\sum_{{\rm even \,} i}\alpha_i\tilde{\bf A}_i)^2\otimes {\bf B}_2^2
\\
&+\frac{1}{4}(\sum_{{\rm odd\,} i}\alpha_i\tilde{\bf A}_i)(\sum_{{\rm even }\, i}\alpha_i\tilde{\bf A}_i)\otimes [{\bf A}_{m,1},{\bf A}_{m,0}]^2
\\
&+\frac{1}{4}(\sum_{{\rm even\,} i}\alpha_i\tilde{\bf A}_i)(\sum_{{\rm odd }\, i}\alpha_i\tilde{\bf A}_i)\otimes [{\bf A}_{m,0},{\bf A}_{m,1}]^2.
\tag{B25}
\end{align*}

In what follows, we assume that $\alpha_1\geq0$ and $\alpha_2\geq0$; Otherwise, by replacing $\alpha_i$ with $-\alpha_i$, the value of the right side of Eq.(B25) is invariable. For commuting Hermitian operators ${\bf A}_{i,j}$ ($\tilde{\bf A}_i, {\bf B}_i$ are commuting) and density operator $\rho$ on Hilbert space $\mathbb{H}_2^m$, we have $[\tilde{\bf A}_{m,i}, \tilde{\bf A}_{m,j}]=0$. From Eq.(B25) we get
\begin{align*}
{\rm Tr}({\cal B}_m^2\rho)
={\rm Tr}((\sum_{{\rm odd\,} i}\alpha_i\tilde{\bf A}_i)^2\otimes{\bf B}_1^2\rho)+
{\rm Tr}((\sum_{{\rm even\,} i}\alpha_i\tilde{\bf A}_i)^2\otimes {\bf B}_2^2\rho).
\tag{B26}
\end{align*}
Two different cases will be discussed as follows:
\begin{itemize}
\item[(i)] $\alpha_{2i-1}\alpha_{2i}\geq0$ for all $i=2, 3, \cdots, 2^{m-1}$. Eq.(B26) implies
\begin{align*}
|{\rm Tr}({\cal B}_m^2\rho)|\leq&\sqrt{(\sum_{{\rm odd \,} i}|\alpha_i|)^2\beta_1^2 +(\sum_{{\rm even \,} i}|\alpha_i|)^2\beta_2^2},
\tag{B27}
\end{align*}
where Eq.(B27) is from the inequalities $\|\tilde{\bf A}_i\|\leq 1$ ($i=1,
\cdots, 2^m$), $\beta^2_{1}=|{\rm Tr}({\bf B}_1^2\rho_m)|=\frac{1}{2}+\frac{1}{4}({\rm Tr}({\bf A}_{m,0}{\bf A}_{m,1}\rho_m)+{\rm Tr}({\bf A}_{m,1}{\bf A}_{m,0}\rho_m))$ and $\beta_{2}^2=|{\rm Tr}({\bf B}_2^2\rho_m)|=\frac{1}{2}-\frac{1}{4}({\rm Tr}({\bf A}_{m,0}{\bf A}_{m,1}\rho_m)+{\rm Tr}({\bf A}_{m,1}{\bf A}_{m,0}\rho_m))$, and $\rho_m$ denotes the reduced state of the $m$-th party. The equality of Eq.(B27) holds for $\rho=\otimes_{i=1}^m\rho_i$, $\tilde{\bf A}_1=\tilde{\bf A}_2=\mathbbm{1}_2$ and $\tilde{\bf A}_j=\frac{\alpha_j}{|\alpha_j|}\mathbbm{1}_2$ when $\alpha_j\not=0$ ($j=3, \cdots, 2^m$). Two subcases will be shown as:
\begin{itemize}
\item[(a)] If $\sum_{{\rm odd \,} i}|\alpha_i|\geq \sum_{{\rm even \,} i}|\alpha_i|$, Eq.(B27) implies that
\begin{align*}
\sup_{\vec{\bf A}^c, \rho}|{\rm Tr}({\cal B}^2\rho)|^{\frac{1}{2}}=\sum_{{\rm odd \,} i}|\alpha_i|,
\tag{B28}
\end{align*}
where the supremum is achieved when $\rho=\frac{1}{2^m}\mathbbm{1}$ and ${\bf A}_{m,0}={\bf A}_{m,1}=\mathbbm{1}_2$. From Schwartz inequality $|{\rm Tr}({\cal B}\rho)|\leq\sqrt{{\rm Tr}({\cal B}^2\rho)}$ we have
\begin{align*}
\sup_{\vec{\bf A}^c,\rho}|{\rm Tr}({\cal B}\rho)| &\leq \sup_{\vec{\bf A}^c,\rho}\sqrt{{\rm Tr}({\cal B}^2\rho)}
\\
&=\sum_{{\rm odd \,} i}|\alpha_i|,
\tag{B29}
\end{align*}
where the equality holds for $\rho=\frac{1}{2^m}\mathbbm{1}$, $\tilde{\bf A}_1=\tilde{\bf A}_2=\mathbbm{1}_2$, and $\tilde{\bf A}_i=\frac{\alpha_i}{|\alpha_i|}\mathbbm{1}_2$ when $\alpha_i\not=0$ ($i=3, \cdots, 2^m$) and ${\bf A}_{m,0}={\bf A}_{m,1}=\mathbbm{1}_2$.

\item[(b)] If $\sum_{{\rm odd \,} i}|\alpha_i|\leq \sum_{{\rm even \,} i}|\alpha_i|$, Eq.(B27) implies that
\begin{align*}
\sup_{\vec{\bf A}^c, \rho}|{\rm Tr}({\cal B}^2\rho)|^{\frac{1}{2}}=\sum_{{\rm even \,} i}|\alpha_i|,
\tag{B30}
\end{align*}
where the supremum is achieved for $\rho=\frac{1}{2^m}\mathbbm{1}$ and ${\bf A}_{m,0}=-{\bf A}_{m,1}=\mathbbm{1}_2$. From Schwartz inequality we have
\begin{align*}
\sup_{\vec{\bf A}^c,\rho}|{\rm Tr}({\cal B}\rho)|&\leq \sup_{\vec{\bf A}^c,\rho}\sqrt{{\rm Tr}({\cal B}^2\rho)}
\\
&=\sum_{{\rm even \,} i}|\alpha_i|,
\tag{B31}
\end{align*}
where the equality holds for $\rho=\frac{1}{2^m}\mathbbm{1}$, $\tilde{\bf A}_1=\tilde{\bf A}_2={\bf A}_{2,0}=-{\bf A}_{2,1}=\mathbbm{1}_2$, and $\tilde{\bf A}_i=\frac{\alpha_i}{|\alpha_i|}\mathbbm{1}_2$ when $\alpha_i\not=0$ ($i=3, \cdots, 2^m$).
\end{itemize}

From Eqs.(B29) and (B31) we obtain
\begin{align*}
\|{\cal B}\|_c=&\sup_{\vec{\bf A}^c,\rho}|{\rm Tr}({\cal B}\rho)|
\\
=&\max\{\sum_{{\rm odd \,} i}|\alpha_i|, \sum_{{\rm even \,} i}|\alpha_i|\}.
\tag{B32}
\end{align*}

\item[(ii)] $\alpha_{2i-1}\alpha_{2i}\geq 0$ for $i\in  {\cal I}_1$ and
$\alpha_{2j-1}\alpha_{2j}\leq 0$ for $j\in  {\cal I}_2$, where ${\cal I}_1$ and ${\cal I}_2$ are partitions of the index set $\{1, \cdots, 2^m\}$. Eq.(B27) implies
\begin{align*}
{\rm Tr}({\cal B}^2\rho)\leq&\max\{(\sum_{{\rm odd \,} i, \atop{ i\in {\cal I}_1}}|\alpha_i|\pm \sum_{{\rm odd \,} j, \atop{  j\in {\cal I}_2}}|\alpha_j| )^2\beta_1^2
+(\sum_{{\rm even \,} k,\atop{ k\in {\cal I}_1}}|\alpha_k|\mp \sum_{{\rm even \,} s, \atop{ s\in {\cal I}_2}}|\alpha_s|)^2\beta_2^2\}
\tag{B33}
\\
\leq &\max\{(\sum_{{\rm odd \,} i,\atop{  i\in {\cal I}_1}}|\alpha_i|\pm \sum_{{\rm odd \,} j,\atop{  j\in {\cal I}_2}}|\alpha_j| )^2,
(\sum_{{\rm even\,} k,\atop{ k\in {\cal I}_1}}|\alpha_k|\mp \sum_{{\rm even \,} s, \atop{ s\in {\cal I}_2}}|\alpha_s|)^2\}
\tag{B34}
\\
=&  \max\{(\sum_{{\rm odd \,} i}|\alpha_i|)^2, (\sum_{{\rm even \,} j}|\alpha_j|)^2\},
\tag{B35}
\end{align*}
where Eq.(B33) is from the inequalities $\|\tilde{\bf A}_l\|\leq 1$ for all indexes $l$; Eq.(B34) is from the equality $\beta_1^2+\beta_2^2=1$ by taking use of similar proofs of Eqs.(B9)-(B12); and $\beta_{i}$ are defined in Eq.(B27). Here, the equality of Eq.(B34) holds for ${\bf A}_{m,0}={\bf A}_{m,1}=\mathbbm{1}_2$ when the right side of Eq.(B33) equals to $(\sum_{{\rm odd \,} i, i\in {\cal I}_1}|\alpha_i|\pm \sum_{{\rm odd \, } j, j\in {\cal I}_2}|\alpha_j| )^2$, or ${\bf A}_{m,0}=-{\bf A}_{m,1}=\mathbbm{1}_2$ when the right side of Eq.(B33) equals to $(\sum_{{\rm even\,} k,{ k\in {\cal I}_1}}|\alpha_k|\mp \sum_{{\rm even \,} s, { s\in {\cal I}_2}}|\alpha_s|)^2$. The equality of Eq.(B33) holds for $\tilde{\bf A}_{2i-1}=\tilde{\bf A}_{2i}=\mathbbm{1}_2$ with $i\in{\cal I}_1$, and $\tilde{\bf A}_{2j-1}=\pm\frac{\alpha_{2j-1}}{|\alpha_{2j-1}|}\mathbbm{1}_2$ ($\alpha_{2j-1}\not=0$) and $\tilde{\bf A}_{2j}=\mp\frac{\alpha_{2j}}{|\alpha_{2j}|}\mathbbm{1}_2$ ($\alpha_{2j}\not=0$) with $j\in {\cal I}_2$. Using Schwartz inequality, similar to Eq.(B29) or (B31) we get the same equality as that given in Eq.(B32).
\end{itemize}

Now, we estimate the norm of ${\cal B}$ in terms of quantum model. For general Hermitian operators ${\bf A}_{i,j}$, from Schwartz inequality and Eq.(B25), we have
\begin{align*}
|{\rm Tr}({\cal B}_m\rho)|
\leq & {\rm Tr}({\cal B}_m^2\rho)^{\frac{1}{2}}
\\
\leq &|{\rm Tr}((\sum_{{\rm odd\,\, } i}\alpha_i\tilde{\bf A}_i)^2\otimes{\bf B}_1^2\rho)
+{\rm Tr}((\sum_{{\rm even \,\,} j}\alpha_j\tilde{\bf A}_j)^2\otimes {\bf B}_2^2\rho)
+{\rm Tr}((\sum_{{\rm odd\,\, } i}\alpha_i\tilde{\bf A}_i)(\sum_{{\rm even \,\,} j}\alpha_j \tilde{\bf A}_j)\rho)
\\
&+{\rm Tr}((\sum_{{\rm even\,\, } j}\alpha_j\tilde{\bf A}_j)(\sum_{{\rm odd \,\,} i}\alpha_i\tilde{\bf A}_i)\rho)|^{\frac{1}{2}}
\tag{B36}
\\
\leq &\sqrt{(\sum_{{\rm odd \,\, } i}|\alpha_i|)^2\beta_1^2
+(\sum_{{\rm even \,\, } i}|\alpha_i|)^2\beta_2^2
+2\sum_{{\rm odd \,\, } i}|\alpha_i|\sum_{{\rm even \,\, } j}|\alpha_j|}
\tag{B37}
\\
\leq &\max\{\sqrt{(\sum_{{\rm odd \,\, } i}|\alpha_i|)^2+2\sum_{{\rm odd \,\, } i}|\alpha_i|\sum_{{\rm even \,\, } j}|\alpha_j|},\sqrt{(\sum_{{\rm even \,\, } i}|\alpha_i|)^2+2\sum_{{\rm odd \,\, } i}|\alpha_i|\sum_{{\rm even \,\, } j}|\alpha_j|}\},
\tag{B38}
\end{align*}
where Eq.(B36) is from the inequalities $\|[\tilde{\bf A}_{i}, \tilde{\bf A}_{j}]\|\leq 2$ for all $i,j$; Eq.(B37) is from the inequalities $\|{\bf A}_{i,j}\|\leq 1$ for all $i,j$; and Eq.(B38) is from the fact $\beta_i^2$ have the same upper bounds for commuting or general Hermitian operators ${\bf A}_{i,j}$, $i=0, 1$. Hence, we obtain
\begin{align*}
\|{\cal B}\|_q=&\sup_{\vec{\bf A}^q, \rho}|{\rm Tr}({\cal B}_m\rho)|
\\
\leq &\max\{\sqrt{(\sum_{{\rm odd \,} i}|\alpha_i|)^2+2\sum_{{\rm odd \, } i}|\alpha_i|\sum_{{\rm even \, } j}|\alpha_j|},\sqrt{(\sum_{{\rm even \,} i}|\alpha_i|)^2+2\sum_{{\rm odd \, } i}|\alpha_i|\sum_{{\rm even \,} j}|\alpha_j|}\}
\\
=&  \max\{\sqrt{3}\sum_{{\rm odd \,} i}|\alpha_i|,\sqrt{3}\sum_{{\rm even \, } j}|\alpha_j|\},
\tag{B39}
\end{align*}
where we have taken use of the inequality $2\sum_{{\rm odd \,} i}|\alpha_i|\sum_{{\rm even \,} j}|\alpha_j|\leq (\sum_{{\rm odd \,} i}|\alpha_i|)^2+(\sum_{{\rm even \,} j}|\alpha_j|)^2$.

From Eqs.(B3), (B32), and (B39), an upper bound of $r_{cq}$ defined in Eq.(B3) is given by
\begin{align*}
r_{cq}\leq &\frac{\max\{\sqrt{3}\sum_{{\rm odd \,} i}|\alpha_i|,\sqrt{3}\sum_{{\rm even \,} j}|\alpha_j|\}}{\max\{\sum_{{\rm odd \, } i}|\alpha_i|,\sum_{{\rm even \,} j}|\alpha_j|\}}
\\
\leq & \sqrt{3}.
\tag{B40}
\end{align*}

In particular, for anticommuting operators (Pauli matrices) ${\bf A}_{i,j}$, we have $\beta_1^2=\beta_2^2=\frac{1}{2}$. Eq.(B33) implies that
\begin{align*}
\|{\cal B}\|_q
\leq &\max\{\sqrt{\frac{1}{2}(\sum_{{\rm odd \,\, } i}|\alpha_i|)^2+2\sum_{{\rm odd \,\, } i}|\alpha_i|\sum_{{\rm even \,\, } j}|\alpha_j|},\sqrt{\frac{1}{2}(\sum_{{\rm even \,\, } i}|\alpha_i|)^2+2\sum_{{\rm odd \,\, } i}|\alpha_i|\sum_{{\rm even \,\, } j}|\alpha_j|}\}
\\
\leq & \max\{\sqrt{\frac{5}{2}}\sum_{{\rm odd \,\, } i}|\alpha_i|,\sqrt{\frac{5}{2}}\sum_{{\rm even \,\, } i}|\alpha_i|\}.
\tag{B41}
\end{align*}
Hence, from Eqs.(B3), (B32), and (B41), the upper bound of $r_{cq}$ is given by
\begin{align*}
r_{cq}\leq \sqrt{\frac{5}{2}}.
\tag{B40}
\end{align*}

\subsection*{Appendix C. Relative critical visibility in the case of general linear Bell inequalities}

Given a general linear Bell operator ${\cal B}=\sum_{i_1,\cdots, i_m}\alpha_{\vec{i}}\hat{\bf A}_{1, i_1}\otimes \cdots \otimes \hat{\bf A}_{m, i_m}$, different from homogeneous Bell operator shown in Eq.(B2), $\hat{\bf A}_{j, i_j}$ may be dichotomic operators ${\bf A}_{j, i_j}$ or excluded. Although one may take use of the identity operator to represent the case of absent observer, we cannot obtain tight upper bound of $\|{\cal B}\|_c$ for all operators ${\cal B}$ using similar method of Appendix B. Here, we provide an accessible method to deal with each general linear Bell operator. A useful decomposition of  ${\cal B}$ is given by
\begin{align*}
{\cal B}=\sum_{j=1}^m{\cal B}_j
\tag{C1}
\end{align*}
where ${\cal B}_j$ are partial Bell operators including measurement operators of the $j$-th party but excluding all operators of $i$-parties for $i=1, \cdots, j-1$, i.e, ${\cal B}_j=\sum_{i_{j},\cdots, i_{m}=0,1}\beta_{i_{j}\cdots i_{m}}{\bf A}_{j,i_j}\hat{\bf A}_{j+1, i_{j+1}}\cdots \hat{\bf A}_{m,i_{m}}$, $\beta_{i_{j}\cdots i_{m}}$ are real coefficients. The main result of this section is the following Lemma.

{\bf Lemma 3}. For ${\cal B}$ presented in Eq.(C1), $r_{cq}$ defined in Eq.(B3) satisfies
\begin{align*}
r_{cq}\leq \sqrt{3}\sum_{i=1}^{m-1}\frac{1}{\gamma_i}+1
\tag{C2}
\end{align*}
if $\|{\cal B}\|_c\geq \gamma_i \|{\cal B}_i\|_c$, $i=1, \cdots, m-1$, where Bell operators ${\cal B}_i$ are defined in Eq.(C1) and $\gamma_i$ are positive constants.

{\bf Proof}. From the decomposition shown in Eq.(C1), ${\cal B}_i$ may be rewritten into
\begin{align*}
{\cal B}_i=\sum_{j=1}^{2^{m+1-i}}\beta_{i,j}\hat{\bf A}_j\otimes{\bf B}_j,
\tag{C3}
\end{align*}
where ${\bf B}_j=\frac{1}{2}({\bf A}_{m+1-i,0}+{\bf A}_{m+1-i,1})$ for all odd integers $j$ satisfying $j\leq 2^{m+1-i}$ and ${\bf B}_j=\frac{1}{2}({\bf A}_{m+1-i,0}-{\bf A}_{m+1-i,1})$ for all even integers $j$ satisfying $j\leq 2^{m+1-k}$, $\beta_{i,j}$ are proper coefficients depending on $\beta_{k_{1}\cdots k_{m+1-i}}$, and $\hat{\bf A}_j$ denote the operators of $\hat{\bf A}_{1, s_1}\otimes \cdots \otimes \hat{\bf A}_{m+2-i, s_{m+2-i}}$ that are Hermitian and satisfy $\|\hat{\bf A}_j\|\leq 1$ and $\hat{\bf A}^2_j=\mathbbm{1}$.

Similar to Eqs.(B24)-(B35), from straight forward computations we obtain
\begin{align*}
\|{\cal B}_k\|_c=&\sup_{\vec{\bf A}^c, \rho}{\rm Tr}({\cal B}_k\rho)
\\
=&\max \{\sum_{{\rm odd \, } i \atop {i\leq N_k}}|\beta_{k,i}|, \sum_{{\rm even \,\, } j \atop {j\leq N_k}}|\beta_{k,j}|\},
\tag{C4}
\end{align*}
where $N_k=2^{m+1-k}$, $k=1, 2, \cdots, m-1$. Moreover, similar to Eqs.(B36)-(B39), we have
\begin{align*}
\| {\cal B}_k\|_c\leq \max\{\sqrt{3}\sum_{{\rm odd \,\, } i \atop {i\leq N_k}}\beta_{k,i},
\sqrt{3}\sum_{{\rm even \,\, } j \atop {j\leq N_k}}\beta_{k,j}\},
\tag{C5}
\end{align*}
where $k=1, 2, \cdots, m-1$. Note that ${\cal B}_m$ is a linear operator including ${\bf A}_{m,0}, {\bf A}_{m,1}$. It follows that
\begin{align*}
&\frac{\sup_{{\bf A}^q_{m,i_m}, \rho_m}{\rm Tr}({\cal B}_m\rho_m)}{\sup_{{\bf A}^c_{m,i_m}, \rho_m}{\rm Tr}({\cal B}_m\rho_m)}= 1,
\tag{C6}
\end{align*}
where $\rho_m$ is the reduced density operator of the $m$-th party.

From Eqs.(C4) and (C6) we have $\sup_{\vec{\bf A}^c, \rho}|{\rm Tr}({\cal B}\rho)|>0$ for any nonzero linear operator ${\cal B}$. In fact, suppose $\sup_{\vec{\bf A}^c, \rho}|{\rm Tr}({\cal B}\rho)|=0$ for all commuting operators ${\bf A}_{i,j}$, ${\cal B}$ must be the zero operator. The reason is that $\sup_{\vec{\bf A}^c, \rho}|{\rm Tr}({\cal B}\rho)|=0$ implies $|{\rm Tr}({\cal B}\rho)|=0$ for all commuting operators ${\bf A}_{i,j}$ and density operator $\rho$. Hence, From Eqs.(C4)-(C6) we have
\begin{align*}
r_{cq}=&\frac{\sup_{\vec{\bf A}^q, \rho}|{\rm Tr}({\cal B}\rho)|}{\sup_{\vec{\bf A}^c, \rho}|{\rm Tr}({\cal B}\rho)|}
\\
\leq &\frac{\sum_{k=1}^{m}\sup_{\vec{\bf A}^q,\hat{\rho}_k}|{\rm Tr}({\cal B}_k\hat{\rho}_k)|}{\sup_{\vec{\bf A}^c, \rho}|{\rm Tr}({\cal B}\rho)|}
\tag{C7}
\\
\leq &\sum_{k=1}^{m}\frac{\sup_{\vec{\bf A}^q, \hat{\rho}_k}|{\rm Tr}({\cal B}_k\hat{\rho}_k)|}{\gamma_k\sup_{\vec{\bf A}^c, \hat{\rho}_k}|{\rm Tr}({\cal B}_k\hat{\rho}_k)|}
\tag{C8}
\\
\leq &\sqrt{3}\sum_{i=1}^{m-1}\frac{1}{\gamma_i}+1,
\tag{C9}
\end{align*}
where $\hat{\rho}_k$ denote the reduced density operators on the subspace determined by ${\cal B}_k$, Eq.(C7) is from the triangle inequality $\sup\{x+y\}\leq \sup\{x\}+\sup\{y\}$, Eq.(C8) is from the assumptions of $\|{\cal B}\|_c\geq  \gamma_i\|{\cal B}_i\|_c$ and $\gamma_i>0$, and Eq.(C9) is from Eqs.(C4)-(C6).

\subsection*{Appendix D. Proof of Theorem 1}

Given an $m$-partite generalized GHZ state, the corresponding Werner state $\rho_v$ is defined in Eq.(A1). We assume $\theta\in (0, \frac{\pi}{2})$. From the equivalent theorem \cite{Deng}, $\rho_v$ is fully separable if and only if $v\leq \frac{\alpha}{2^m+\alpha}$ with $\alpha=\frac{2}{\sin2\theta}$. It follows that the critical parameter $v^*$ for which Werner states $\rho_v$ with $v\leq v^*$ are fully separable is given by
\begin{align*}
v^*=\frac{1}{2^{m-1}\sin 2\theta+1}.
\tag{D1}
\end{align*}

In what follows, we prove that the critical visibility $v'$ for which Werner state violates some linear Bell inequality is strictly larger than $v^*$ presented in Eq.(D1). Assume that an arbitrary $m$-partite Bell inequality is given by
\begin{align*}
\|{\cal B}\|_{c(q)}\leq c_1(c_2),
\tag{D2}
\end{align*}
where $c_1$ and $c_2$ denote the respective upper bound of the norm of Bell operator ${\cal B}$ in the case of commuting Hermitian measurement operators and general Hermitian measurement operators of all observers. From Lemma 1, the critical visibility $v'$ for which Werner state defined in Eq.(A1) violates Bell inequality of Eq.(D2) with $c_1$ is given by
\begin{align*}
v'\geq \frac{2^m-1}{2^mr_{cq}-1},
\tag{D3}
\end{align*}
where $r_{cq}$ is defined in Eq.(B3).

Case 1. Homogeneous linear Bell inequalities.

From Lemma 2 and Eq.(D3), we get $v'\geq \frac{2^m-1}{2^m\sqrt{3}-1}$. When $\sin2\theta>\frac{2\sqrt{3}-2}{2^m-1}$ we have $v'\geq\frac{2^m-1}{2^m\sqrt{3}-1}>\frac{1}{2^{m-1}\sin2\theta+1}=v^*$ from Eq.(D1). So, Werner states shown in Eq.(A1) cannot be completely detected by homogeneous linear Bell inequalities when $\theta$ satisfies
\begin{align*}
\frac{1}{2}\arcsin(\frac{2\sqrt{3}-2}{2^m-1})
<\theta<\pi-\frac{1}{2}\arcsin(\frac{2\sqrt{3}-2}{2^m-1}).
\tag{D4}
\end{align*}
Numerical bounds of $\theta$ presented in Eq.(D4) are shown in Table I. It follows that for almost all generalized GHZ states, the corresponding Werner states cannot be completely detected by homogeneous linear Bell inequalities.

\begin{table}[htbp]
\centering
\caption{Undetectable Werner states of $m$-partite generalized GHZ states defined in Eq.(A1) in the case of homogeneous linear Bell inequalities. $\theta_u$ denotes the upper bound of $\theta$. $\theta_l$ denotes the lower bound of $\theta$. $\theta_u$ and $\theta_l$ are given in Eq.(D4). $r$ is defined by $r=\frac{1}{\pi}(\theta_u-\theta_l)$, which denotes the measure of generalized GHZ state (in the set of all generalized GHZ states) whose Werner states cannot be completely detected.}
\begin{tabular}{c|ccccc}
   \hline
$m$  & 2 & 3 & 4 & 5 & 6
\\ \hline
${\theta_u}/{\pi}$ & 0.9189  &  0.9665  &  0.9844  &  0.9925 & 0.9963
\\
${\theta_l}/{\pi}$  & 0.0811 &  0.0335  &  0.0156  &  0.0075 & 0.0037
\\
$r$ & 83.77\% & 83.29\% & 96.89\% & 98.5\% & 99.26\%
\\
   \hline
\end{tabular}
\end{table}

Case 2. General linear Bell inequalities.

Denote $\gamma=\sqrt{3}\sum_{i=1}^{m-1}\gamma_i^{-1}+1$. From Lemma 3 and Eq.(D3), we obtain $v'\geq \frac{2^m-1}{2^m\gamma-1}$ when $\|{\cal B}\|_c\geq \gamma_i\|{\cal B}_i|_c$, $i=1, \cdots, m-1$. From Eq.(D1), when $\sin2\theta>\frac{2\sqrt{3}}{2^m-1}\sum_{i=1}^{m-1}\frac{1}{\gamma_i}$ we have $v'\geq \frac{2^m-1}{2^m\gamma-1}>\frac{1}{2^{m-1}\sin2\theta+1}=v^*$. Hence, there are Werner states defined in Eq.(A1) that cannot be completely detected by a given linear Bell inequality shown in Eq.(D2) when $\theta$ satisfies
\begin{align*}
\frac{1}{2}\arcsin(\frac{2\sqrt{3}}{2^m-1}\sum_{i=1}^{m-1}\frac{1}{\gamma_i})
<\theta
<\pi-\frac{1}{2}\arcsin(\frac{2\sqrt{3}}{2^m-1}\sum_{i=1}^{m-1}\frac{1}{\gamma_i}).
\tag{D5}
\end{align*}
It implies that Werner states of generalized GHZ states with large $m$ cannot be completely detected by linear Bell inequalities with all $\gamma_i$s satisfying $\gamma_i=\frac{1}{{\rm Poly}(m)}$, i.e, polynomial functions of $m$.

In order to complete the proof, we need to estimate the bounds of $\gamma_i$. Unfortunately, we cannot get explicit bounds for all $\gamma_i$s. Here, we take use of numeric methods. For a general linear Bell operator ${\cal B}=\sum_{i=1}^{m-1}{\cal B}_i$ given in Eq.(C1), the classical bound is always achieved from extremal points (i.e., ${\bf A}_{i,j}=\pm\mathbbm{1}_2$) when $\|{\bf A}_{i,j}\|\leq 1$ \cite{Tsi1,Tsi2}. In this case, all correlations $\langle \hat{\bf A}_{1, i_1}\cdots \hat{\bf A}_{m,i_{m}}\rangle$ of $\langle{\cal B}\rangle$ equal to $\prod_{j=1}^m\langle \hat{\bf A}_{j, i_j}\rangle$. From Eq.(C1), for each Bell operator ${\cal B}$, $\langle{\cal B}\rangle$ may be vectorized as $\vec{\alpha}\cdot \vec{A}$ and $\langle{\cal B}_j\rangle$ may be vectorized as $\vec{\alpha}_j\cdot \vec{A}_j$, where $\vec{\alpha}$ denotes $3^m-1$-dimensional vector consisting of all coefficients $\beta_{i_{1}\cdots i_{m}}$ of ${\cal B}$, $\vec{A}$ denotes $3^m-1$-dimensional vector consisting of all correlations $\prod_{j=1}^m\langle \hat{\bf A}_{j, i_j}\rangle$ of $\langle {\cal B}\rangle$, $\vec{\alpha}_j$ denotes $\ell_j$-dimensional vector consisting of all coefficients $\beta_{i_{j}\cdots i_{m}}$ of ${\cal B}_j$, $\vec{A}_j$ denotes $\ell_j$-dimensional vector consisting of all correlations $\prod_{s=j}^m\langle {\bf A}_{s, i_s}\rangle$ of $\langle{\cal B}_j\rangle$, and $\ell_j=2\times 3^{m-j}$, $j=1, \cdots, m$. From Eqs.(C9) and (D5), it is sufficient to get the tight bound of the relative critical visibility $r_{cq}$ by estimating $\min_{\cal B}\gamma_j$ from $\min_{\cal B}\frac{\|{\cal B}\|_c}{\|{\cal B}_j\|_c}$. Here, we only need to consider all coefficients of ${\cal B}$ on $3^m-1$-dimensional unit hypersphere. Denote $M=[\vec{\bf A}]$ as a $4^{m}\times (3^m-1)$ matrix and $M_j=[\vec{\bf A}_j]$ as a $4^{m+1-j}\times \ell_j$ matrix, where $\langle{\bf A}_{j, i_j}\rangle, \cdots, \langle{\bf A}_{m, i_m}\rangle\in\{\pm 1\}$, $j=1, \cdots, m$. We present Algorithm 1.

\begin{algorithm}[h]
\caption{Evaluating $\min_{\cal B}\gamma_i$ in the case of $m$-partite Bell inequalities}
\KwIn{$m$, $N$}
\KwOut{$\min_{\cal B}{\gamma}_1, \cdots, \min_{\cal B}{\gamma}_m$}
\begin{itemize}
\item[]${\Xi}$ is an $N$-partition of $3^{m}-1$-dimensional (column) sphere;

\item[]{\bf For $i=1:m$}

\item[] \quad $X_i={\Xi}(:,L_{i-1}:L_{i})$;

\item[]\quad{\bf For $j=1:N$}

\item[] \qquad  $\vec{x}_{ij}=X_i(:,j)$;

\item[] \qquad $\|\hat{\cal B}_j\|_c=\max |M\vec{x}_{ij}|$;

\item[] \qquad $\|\hat{\cal B}_{ij}\|_c=\max |M_i\vec{x}_{ij}(L_{i-1}:L_{i})|$;

\item[] \qquad $\gamma_{ij}=\frac{\|\hat{\cal B}_j\|_c}{\|\hat{\cal B}_{ij}\|_c}$;

\item[]\quad ${\gamma_i}=\min_j\{\gamma_{ij}\}$;
\end{itemize}
\end{algorithm}

In Algorithm 1, $N$ denotes the total number of unit vectors, where each unit vector can determine a general $m$-partite Bell operator. ${\Xi}$ is a $(3^{m}-1)\times L_{m}$ matrix, where each column vector is a $3^{m}-1$-dimensional real unit vector, $L_i=\sum_{j=1}^{i-1}\ell_j$ and $L_{-1}=0$. $N$-partition of unit sphere means that there are $N$ points on unit sphere such that the minimum distance of any two points has the lower bound $O(N^{1/m})$. $X_i={\Xi}(:,L_{i-1}:L_{i})$ denotes the submatrix of ${\Xi}$, which includes all components from the $L_{i-1}$-th column to the $L_{i}$-th column, $i=1, \cdots, m$. $\vec{x}_{ij}$ denotes the $j$-th column of the matrix $X_i$. ${\cal B}_j$ and ${\cal C}_{ij}$ are the corresponding Bell operators derived from coefficients $\vec{x}_{ij}$. $\| \hat{\cal B}_j\|_c$ and $\| \hat{\cal B}_{ij}\|_c$ are computed as the maximal bound of extremal points. Note that $\frac{\|\hat{\cal B}_j\|_c}{\|\hat{\cal B}_{ij}\|_c}$ is continue function in terms of variables $\vec{\alpha}_j$. Algorithm 1 provides a useful approximate algorithm to obtain lower bound $\gamma_i$. The computation complexity is $O(mN^23^m2^{5m+m^2}\prod_{i=1}^m\ell_i)$ that is exponential in $m$. Numeric evaluations of small $m$ are shown in Table II. From Table II, all $\gamma_i$s satisfy $\gamma_i\geq 1$. Numeric bounds of $\theta$ given by Eq.(D5) for $\gamma_1=\cdots =\gamma_m=1$ are shown in Table III. Although there are Werner states that cannot be detected for $m=2$ \cite{ZB}, however, from Table III there is no example. The reason is that the quantum bound shown in Eq.(B40) is not tight.

\begin{table}[htbp]
\centering
\caption{The lower bounds of $\gamma_i$s in the case of $m$-partite Bell inequalities. All lower bounds are larger than 1. We conjecture that all lower bounds of $\gamma_i$ for each general Bell inequality are larger than $\frac{1}{{\rm Poly}(m)}$ for some polynomial function ${\rm Poly}(m)$.}
\begin{tabular}{c|cccccccccc}
   \hline
$m$  & $\gamma_1$ & $\gamma_2$ & $\gamma_3$ & $\gamma_4$ & $\gamma_5$ & $\gamma_6$
\\ \hline
$2$ & 1.00  &  1.01
\\
$3$ & 1.00  &  1.10 & 1.72
\\
$4$ & 1.00  &  1.10 & 1.80  & 3.10
\\
$5$ & 1.00  &  1.20 & 1.90  & 3.40  & 6.70
\\
$6$ & 1.00  &  1.23 & 2.00 & 3.95  & 6.98  &  12.00
\\
   \hline
\end{tabular}
\end{table}

\begin{table}[htbp]
\centering
\caption{Numeric bounds of $\theta$ given in Eq.(D5). Here, we assume $\gamma_1=\cdots=\gamma_m=1$. $\theta_u$ denotes the upper bound of $\theta$ given in Eq.(D5). $\theta_l$ denotes the lower bound of $\theta$. $r$ is defined by $r =\frac{1}{\pi}(\theta_u-\theta_l)$, which denotes the measure of generalized GHZ states (in the set of all generalized GHZ states) whose Werner states cannot be completely detected.}
\begin{tabular}{c|ccccccccccccc}
\hline
$m$ & 2 & 3 & 4 &  5 & 6
\\ \hline
${\theta_l}/{\pi}$ & - &0.2272 & 0.1218 & 0.0738 & 0.0443
\\
${\theta_u}/{\pi}$ & - & 0.7728 & 0.8782 & 0.9262 & 0.9557
\\ \hline
$r$ & - & 54.56\% & 75.64\% & 85.24\% & 91.14\%
\\
   \hline
\end{tabular}
\end{table}

\section*{Appendix E. Critical parameters of Werner states}

In this section, we take use of the notations of index vector defined in Appendix A. Given an $m$-partite pure state $|\Phi\rangle=\sum_{\vec{i}}\alpha_{\vec{i}\,}|\vec{i}\,\rangle$, the corresponding Werner state is defined by
\begin{align*}
\rho_v=\frac{1-v}{2^m}\mathbbm{1} +v|\Phi\rangle\langle \Phi|,
\tag{E1}
\end{align*}
where $\mathbbm{1}$ denotes the identity operator on Hilbert space $\mathbb{H}^{\otimes m}_2$, $\alpha_{\vec{i}\,}$ are complex coefficients satisfying $\sum_{\vec{i}\,}|\alpha_{\vec{i}\,}|^2=1$ and $v\in [0, 1]$. It easily proves that $\rho_v$ are density matrices for all $v\in[0,1]$.

The proof of Theorem 2 is completed within the following three sections. Appendices E and F contribute to the upper bound of the critical parameter $v^*$ for which all Werner states $\rho_v$ defined in Eq.(E1) with $v\leq v^*$ are fully separable. Appendix G provides the proof of Theorem 2. In this section, we present a parameter-dependent upper bound of $v^*$ as follows:

{\bf Lemma 4}. The critical parameter $v^*$ for which Werner states defined in Eq.(E1) with $v\leq v^*$ are fully separable satisfies
\begin{align*}
v^*\leq& \min\{\min_{\vec{i}\in {\cal I},\vec{j}}\frac{1}{\sqrt{|4^{m}|\alpha_{\vec{j}}|^2|\alpha_{\vec{j}_c}|^2-4^{m}
|\alpha_{\vec{i}}|^2|\alpha_{\vec{i}_c}|^2
+2^m(|\alpha_{\vec{i}}|^2+|\alpha_{\vec{i}_c}|^2)-1|}},
1\}
\tag{E2}
\end{align*}
where ${\cal I}$ denotes the index set
${\cal I}=\{\vec{i}\, ||\alpha_{\vec{i}}|^2+|\alpha_{\vec{i}_c}|^2\leq \frac{1}{2^{m-1}}\}$ and the index vector $\vec{j}$ can be any $m$-bit vector.

{\bf Proof}. The proof is to obtain the necessary condition of Werner states being fully separable states. From \cite{Pitt}, the necessary condition that $\rho_v$ in Eq.(E1) is fully separable is given by
\begin{align*}
\min_{\vec{i}}\sqrt{\langle \vec{i}|\rho_v|\vec{i}\,\rangle\langle \vec{i}_c|\rho_v|\vec{i}_c\rangle}\geq\max_{\vec{j}}|\langle\vec{j}|\rho|\vec{j}_c\rangle|.
\tag{E3}
\end{align*}

In what follows, we estimate the upper bound of $v^*$ from Eq.(E3). From Eq.(E1), we have
\begin{align*}
\langle\vec{j}|\rho_v|\vec{j}_c\rangle^2-\langle \vec{i}|\rho_v|\vec{i}\,\rangle\langle \vec{i}_c|\rho_v|\vec{i}_c\rangle
=&v^2|\alpha_{\vec{j}}|^2|\alpha_{\vec{j}_c}|^2
-(\frac{1-v}{2^m}+v|\alpha_{\vec{i}}|^2)(\frac{1-v}{2^m}
+v|\alpha_{\vec{i}_c}|^2)
\\
=&\frac{1}{4^m}[(4^m|\alpha_{\vec{j}}|^2|\alpha_{\vec{j}_c}|^2-
(2^m|\alpha_{\vec{i}}|^2-1)(2^m|\alpha_{\vec{i}_c}|^2-1))v^2
\\
&-(2^m|\alpha_{\vec{i}}|^2+2^m|\alpha_{\vec{i}_c}|^2-2)v-1]
\\
\leq&0
\tag{E4}
\end{align*}
for all $m$-bit index vectors $\vec{i}, \vec{j}$.

Denote two functions $f(\vec{i},\vec{j}\,)$ and $g(\vec{i}\,)$ as
 \begin{align*}
f(\vec{i},\vec{j}\,)=&4^{m}|\alpha_{\vec{j}}|^2|\alpha_{\vec{j}_c}|^2-4^{m}
|\alpha_{\vec{i}}|^2|\alpha_{\vec{i}_c}|^2
+2^m(|\alpha_{\vec{i}}|^2+|\alpha_{\vec{i}_c}|^2)-1,
\tag{E5}
\\
g(\vec{i}\,)=&2^m|\alpha_{\vec{i}}|^2+2^m|\alpha_{\vec{i}_c}|^2-2.
\tag{E6}
\end{align*}
From the normalization condition of $\sum_{\vec{i}}(|\alpha_{\vec{i}}|^2+|\alpha_{\vec{i}_c}|^2)=1$, we get $|\alpha_{\vec{i}}|^2+|\alpha_{\vec{i}_c}|^2\leq \frac{1}{2^{m-1}}$ for some index vector $\vec{i}$ using the pigeonhole principle because there are $2^{m-1}$ index vectors $\vec{i}$. So, from Eq.(E6) we obtain
\begin{align*}
\min_{\vec{i}}g(\vec{i}\,)=\min_{\vec{i}\in {\cal I}}g(\vec{i}\,)\leq 0,
\tag{E7}
\end{align*}
where ${\cal I}$ denotes the index set of
${\cal I}=\{\vec{i}\,|g(\vec{i}\,)\leq 0\}$. The following proof is divided into three cases.

Case 1. $f(\vec{i},\vec{j}\,)> 0$ for some index vectors $\vec{i}\in {\cal I}$ and $\vec{j}$. Eq.(E4) implies that
\begin{align*}
v \leq \frac{g(\vec{i}\,)+\sqrt{\Delta(\vec{i},\vec{j}\,)}}{2f(\vec{i},\vec{j}\,)},
\tag{E8}
\end{align*}
where $\Delta(\vec{i},\vec{j}\,)=g(\vec{i}\,)^2+4f(\vec{i},\vec{j}\,)$. Since $\sqrt{\Delta(\vec{i},\vec{j}\,)}\leq |g(\vec{i}\,)|+2\sqrt{f(\vec{i},\vec{j}\,)}$, Eq.(E8) yields
to
\begin{align*}
v \leq \sqrt{\frac{g(\vec{i}\,)+|g(\vec{i}\,)|}{2f(\vec{i},\vec{j}\,)}+
\frac{1}{f(\vec{i},\vec{j}\,)}}.
\tag{E9}
\end{align*}
Note that Eq.(E4) holds for any index vectors $\vec{i},\vec{j}$. From Eqs.(E7) and (E9) we have
\begin{align*}
v \leq & \min_{\vec{i},\vec{j}}\sqrt{\frac{g(\vec{i}\,)+|g(\vec{i}\,)|}{2f(\vec{i},\vec{j}\,)}+
\frac{1}{f(\vec{i},\vec{j}\,)}}
\\
\leq &\min_{\vec{i}\in {\cal I},\vec{j}}\frac{1}{\sqrt{f(\vec{i},\vec{j}\,)}},
\tag{E10}
\end{align*}
where ${\cal I}$ is defined in Eq.(E7).

Case 2. $f(\vec{i},\vec{j}\,)<0$ and $\Delta(\vec{i},\vec{j}\,)\geq0$ for some index vectors $\vec{i}\in {\cal I}$ and $\vec{j}$. Eq.(E4) implies that
\begin{align*}
v \leq \frac{g(\vec{i}\,)+\sqrt{\Delta(\vec{i},\vec{j}\,)}}{2f(\vec{i},\vec{j}\,)}
\tag{E11}
\end{align*}
or
\begin{align*}
v \geq \frac{g(\vec{i}\,)-\sqrt{\Delta(\vec{i},\vec{j}\,)}}{2f(\vec{i},\vec{j}\,)},
\tag{E12}
\end{align*}
where $\Delta(\vec{i},\vec{j}\,)$ is defined in Eq.(E8). Since $|g(\vec{i}\,)|-2\sqrt{-f(\vec{i},\vec{j}\,)}\leq \sqrt{\Delta(\vec{i},\vec{j}\,)}\leq |g(\vec{i}\,)|$, from Eq.(E11) we obtain
\begin{align*}
v \leq &\frac{-g(\vec{i}\,)-\sqrt{\Delta(\vec{i},\vec{j}\,)}}{-2f(\vec{i},\vec{j}\,)}
\\
\leq &\frac{-g(\vec{i}\,)+2\sqrt{-f(\vec{i},\vec{j}\,)}
-|g(\vec{i}\,)|}{-2f(\vec{i},\vec{j}\,)}
\tag{E13}
\end{align*}
for any index vectors $\vec{i},\vec{j}$ because Eq.(E4) holds for any index vectors $\vec{i},\vec{j}$.

Note that Eq.(E13) implies
\begin{align*}
v\leq&\min_{\vec{i},\vec{j}}
\frac{-g(\vec{i}\,)+2\sqrt{-f(\vec{i},\vec{j}\,)}-|g(\vec{i}\,)|}{-2f(\vec{i},\vec{j}\,)}
\\
\leq & \min_{\vec{i}\in {\cal I},\vec{j}} \frac{1}{\sqrt{-f(\vec{i},\vec{j}\,)}}.
\tag{E14}
\end{align*}
Moreover, from Eq.(E12) we get
\begin{align*}
v \geq &\frac{g(\vec{i}\,)+|g(\vec{i}\,)|}{-2f(\vec{i},\vec{j}\,)}
\\
\geq &\max_{\vec{i}} \frac{g(\vec{i}\,)+|g(\vec{i}\,)|}{-2f(\vec{i},\vec{j}\,)}
\\
\geq &0.
\tag{E15}
\end{align*}

Case 3. $f(\vec{i},\vec{j}\,)<0$ and $\Delta(\vec{i},\vec{j}\,)<0$ for some index vectors $\vec{i}\in {\cal I}$ and $\vec{j}$. For this case, Eq.(E4) implies $v\in [0,1]$.

To sum up, since Eq.(E4) holds for all index vectors $\vec{i}, \vec{j}$,  from Eqs.(E10) and (E14) we have
\begin{align*}
v \leq \min\{\min_{\vec{i}\in {\cal I},\vec{j}} \frac{1}{\sqrt{|f(\vec{i},\vec{j}\,)|}},1\}
\tag{E16}
\end{align*}
Here, we do not consider the case of $f(\vec{i},\vec{j}\,)=0$ for some index vectors $\vec{i}\in {\cal I}$ and $\vec{j}$, which yields to $v\leq -1/g(\vec{i}\,)$. The reason is from $f(\vec{i},\vec{j}\,)\not=0$ for almost all pure states. Of course, the minimal upper bound of $v^*$ derived from Eq.(E4) is lower than the bound presented in Eq.(E16). Hence, we have completed the proof.

\section*{Appendix F. Special upper bound of critical parameters}

In this section, we present a special upper bound of critical parameters which is independent of coefficients of pure states. Note that the separability of each pure is independent of global phases. We only need to consider all pure states which consist of a unit hypersphere (embedded in a higher-dimensional Euclidean space).

{\bf Lemma 5}. There is a nonzero measure of pure states $|\Phi\rangle=\sum_{i_1, \cdots, i_m=0, 1}\alpha_{\vec{i}\,}|\vec{i}\,\rangle$ satisfying
\begin{align*}
\max_{\vec{i}\in {\cal I},\vec{j}}\sqrt{|f(\vec{i},\vec{j}\,)|}
>\frac{2^{m}{\rm Poly}(m)-1}{2^m-1}
\tag{F1}
\end{align*}
for large $m$, where nonzero measure means the ratio of Lebesgue measures of pure states satisfying Eq.(F1) and all pure states is nonzero, $f(\vec{i},\vec{j}\,)$ is defined in Eq.(E5), ${\rm Poly}(m)$ denotes any polynomial function of $m$, and ${\cal I}$ is defined in Eq.(E2).

{\bf Proof}. Denote $\delta=\frac{2^{m}{\rm Poly}(m)-1}{2^m-1}$. Define two index sets ${\cal J}_1$ and ${\cal J}_2$ as
\begin{align*}
{\cal J}_1:=\{(\vec{i}, \vec{j}\,)|f(\vec{i},\vec{j}\,)\geq 0, \vec{i}\in {\cal I} \},
\tag{F2}
\\
{\cal J}_2:=\{(\vec{i}, \vec{j}\,)|f(\vec{i},\vec{j}\,)\leq 0, \vec{i}\in {\cal I} \}.
\tag{F3}
\end{align*}
Since $\max_{\vec{i}\in {\cal I}}\{|\alpha_{\vec{i}}|^2|\alpha_{\vec{i}_c}|^2-\frac{1}{2^m}
(|\alpha_{\vec{i}}|^2+|\alpha_{\vec{i}_c}|^2)\}\leq 0$, we obtain
\begin{align*}
\max_{(\vec{i}, \vec{j}\,)\in {\cal J}_2}|f(\vec{i},\vec{j}\,)|=&
\max_{(\vec{i}, \vec{j}\,)\in {\cal J}_1}\{1-4^{m}|\alpha_{\vec{j}}|^2|\alpha_{\vec{j}_c}|^2+4^{m}(
|\alpha_{\vec{i}}|^2|\alpha_{\vec{i}_c}|^2-\frac{1}{2^m}(|\alpha_{\vec{i}}|^2
+|\alpha_{\vec{i}_c}|^2)\}
\\
\leq &\max_{(\vec{i}, \vec{j}\,)\in {\cal J}_2}\{1-4^{m}|\alpha_{\vec{j}}|^2|\alpha_{\vec{j}_c}|^2\}
\\
\leq & 1.
\tag{F4}
\end{align*}
Eq.(F4) means that there is no index pair $(\vec{i},\vec{j}\,)\in {\cal J}_2$ satisfying the inequality shown in Eq.(F1) because $\delta>1$ from its definition.

Now, assume that
\begin{align*}
&\max_{(\vec{i}, \vec{j}\,)\in {\cal J}_1}|f(\vec{i}, \vec{j}\,)|=
\max_{(\vec{i}, \vec{j}\,)\in {\cal J}_1}f(\vec{i}, \vec{j}\,)> \delta^2.
\tag{F5}
\end{align*}
Since $\max_{\vec{i}\in {\cal I}}(|\alpha_{\vec{i}}|^2|\alpha_{\vec{i}_c}|^2
-\frac{1}{2^m}(|\alpha_{\vec{i}}|^2+|\alpha_{\vec{i}_c}|^2))\leq 0$, Eq.(F5) holds when the following inequality is satisfied
\begin{align*}
\max_{\vec{j}}|\alpha_{\vec{j}}|^2|\alpha_{\vec{j}_c}|^2
>\frac{({\rm Poly}(m)+1)^2}{4^{m}}>\frac{\delta^2}{4^{m}}.
\tag{F6}
\end{align*}

Define $\mathbb{S}$ as the set of all normalized $m$-partite pure states and $\mathbb{S}_m$ as the set of all normalized $m$-partite pure states satisfying Eq.(F6), i.e.,
\begin{align*}
\mathbb{S}_m:=\{|\Phi\rangle|\max_{\vec{j}}|\alpha_{\vec{j}}|^2|\alpha_{\vec{j}_c}|^2
>\frac{({\rm Poly}(m)+1)^2}{4^{m}}, \sum_{\vec{j}}|\alpha_{\vec{j}}|^2=1
 \}.
\tag{F7}
\end{align*}

Note that $(\vec{i}, \vec{j}\,) \in{\cal J}_1$ when Eq.(F7) is satisfied. It means that all states in $\mathbb{S}_m$ satisfy Eq.(F1). Since $\max_{\vec{j}}|\alpha_{\vec{j}}|^2|\alpha_{\vec{j}_c}|^2\in (0, \frac{1}{4}]$, one may informally conclude that the most of pure states $|\Phi\rangle$ satisfy Eq.(F1) for large integer $m$, where the upper bound $\frac{1}{4}$ is achieved for the maximally entangled $m$-partite GHZ state.

We may formally prove the result as follows. Denote $\overline{\mathbb{S}}_m:=\{|\Phi\rangle|\max_{\vec{j}}|\alpha_{\vec{j}}
|^2|\alpha_{\vec{j}_c}|^2\leq 4^{-m}({\rm Poly}(m)+1)^2,\sum_{\vec{j}}|\alpha_{\vec{j}}|^2=1\}$, i.e., all pure states that do not satisfy Eq.(F1). By representing each coefficient $\alpha_{\vec{j}}$ by $\alpha_{\vec{j}}=r_{\vec{j\,}}e^{\sqrt{-1}\theta_{\vec{j\,}}}$, $\overline{\mathbb{S}}_m$ and $\overline{\mathbb{S}}$ are decomposed into product spaces of $\overline{\mathbb{S}}_m=\mathbb{S}^r_m\times [0, 2\pi]^{\times 2^m}$ and $\mathbb{S}=\mathbb{S}^r\times [0, 2\pi]^{\times 2^m}$, where $\overline{\mathbb{S}}^r_m:=\{|\Phi\rangle|\max_{\vec{j}}r_{\vec{j}}^2
r_{\vec{j}_c}^2 \leq 4^{-m}({\rm Poly}(m)+1)^2,\sum_{\vec{j}}r_{\vec{j}}^2=1, r_{\vec{j}}\geq0\}$,
$\mathbb{S}^r_m:=\{|\Phi\rangle|\sum_{\vec{j}}r_{\vec{j}}^2=1, r_{\vec{j}}\geq0\}$, and $[0, 2\pi]^{\times 2^m}$ denotes $2^m$-dimensional phase space of $(\theta_{\vec{0}}, \cdots, \theta_{\vec{1}})$. Hence, we only need to consider subspaces $\mathbb{S}^r$ and $\overline{\mathbb{S}}^r_m$ which are independent of the phase space. Specially, we have
\begin{align*}
\frac{M(\mathbb{S}_m)}{M(\mathbb{S})}&=1
-\frac{M(\overline{\mathbb{S}}_m)}{M(\mathbb{S})}
\\
&=1
-\frac{M(\overline{\mathbb{S}}^r_m)}{M(\mathbb{S}^r)},
\tag{F8}
\end{align*}
where $M(\cdot)$ denotes Lebesgue measure on a $2^m$-dimensional real hypersphere (embedded in a $2^m+1$-dimensional Euclidean space).

Denote $c=2^{-m}({\rm Poly}(m)+1)$, for all states in $\overline{\mathbb{S}}^r_m$, we have $r_1r_{2^m}\leq c$. Denote $\mathbb{S}_+^{2^m}$ as the subset of $2^m$-dimensional unit hypersphere $\mathbb{S}^{2^m}$ with all positive components. Note that the measure of $2^{m}$-dimensional hypersphere $\mathbb{S}^{2^m}$ with radius $x$ is given by $M(\mathbb{S}^{2^m})=\frac{2\pi^{2^{m-1}}x^{2^m}}{\Gamma(2^{m-1})}$, where Gamma function $\Gamma(k)=k!$. From the symmetry of hypersphere we get
\begin{align*}
M(\mathbb{S}^r)=&M(\mathbb{S}_+^{2^m})
\\
=&\frac{1}{2^{2^m}}M(\mathbb{S}^{2^m})
\\
=&\frac{2\pi^{2^{m-1}}}{2^{2^m}\Gamma(2^{m-1})}.
\tag{F9}
\end{align*}
For set $\overline{\mathbb{S}}_m^r$ we have
\begin{align*}
M(\overline{\mathbb{S}}_m^r)
\leq &\int_{\vec{r}\in S_+^{2^m}|r_1r_{2^m}\leq c}dx
\\
=&\frac{2\pi^{2^{m-1}-1}}{2^{2^m}\Gamma(2^{m-1}-1)}\int_{r_1r_{2^m}\leq c}(1-r_1^2
-r_{2^m}^2)^{2^{m-1}-1}dr_1dr_{2^m}.
\tag{F10}
\end{align*}

Let $r_1=r\cos \theta, r_{2^m}=r\sin\theta$, $r\in[0, c]$ and $\theta\in[0,\frac{\pi}{2}]$. From Eq.(F10) we get
\begin{align*}
M(\overline{\mathbb{S}}_m^r)
\leq & \frac{2\pi^{2^{m-1}-1}}{2^{2^m}\Gamma(2^{m-1}-1)}\int_{r\in[0, c], \theta\in[0,\frac{\pi}{2}]|r\sin2\theta\leq 2c}r(1-r^2)^{2^{m-1}-1}drd\theta
\\
=&\frac{2\pi^{2^{m-1}-1}}{2^{2^m}\Gamma(2^{m-1}-1)}\int_{r\in[0, c]}\int_{\theta\in [0,\frac{1}{2}\arcsin\frac{2c}{r}]}r(1-r^2)^{2^{m-1}-1}drd\theta
\\
\leq & \frac{\pi^{2^{m-1}}}{2^{2^m}\Gamma(2^{m-1}-1)}\int_{r\in[0, c]}r(1-r^2)^{2^{m-1}-1}dr
\\
\leq & \frac{2\pi^{2^{m-1}}}{2^{2^m}\Gamma(2^{m-1})}- \frac{\pi^{2^{m-1}}}{2^{2^m}\Gamma(2^{m-1})}(1-c^2)^{2^{m-1}}.
\tag{F11}
\end{align*}

Consequently, Eqs.(F8)-(F11) implies that
\begin{align*}
\frac{M(\mathbb{S}_m)}{M(\mathbb{S})}&\geq (1-c^2)^{2^{m-1}}.
\tag{F12}
\end{align*}
It means that there is a nonzero measure of pure states satisfying Eq.(F1).

\section*{Appendix G. Proof of Theorem 2}

Given an $m$-partite pure state $|\Phi\rangle=\sum_{\vec{i}}\alpha_{\vec{i}}|\vec{i}\,\rangle$, the corresponding Werner state is defined in Eq.(E1). In what follows, we prove that the critical visibility $v'$, for which $\rho_v$ with $v>v'$ violates some linear Bell inequality, is strictly larger than $v^*$ given in Eq.(F1).

Assume an $m$-partite Bell inequality is given by
\begin{align*}
\| {\cal B}\|\leq c_1(c_2),
\tag{G1}
\end{align*}
where $c_1$ and $c_2$ denote the respective classical bound (in terms of LHS model) and quantum bound (in terms of quantum model). From Lemma 1, the critical visibility $v'$ of Werner state $\rho_v$ defined in Eq.(E1) is given by
\begin{align*}
v'\geq \frac{2^m-1}{2^mr_{cq}-1}
\tag{G2}
\end{align*}
for which Werner state violates linear Bell inequality presented in Eq.(G1) with $c_1$, where $r_{cq}$ is defined in Eq.(B3). The following proof will be divided into two cases.

Case 1. Homogeneous linear Bell inequalities. From Lemma 2 and Eq.(G2), we have $v'\geq \frac{2^m-1}{2^m\sqrt{3}-1}$. From Lemma 5, there is a nonzero measure of pure states satisfying $\max_{\vec{i}\in {\cal I},\vec{j}}\sqrt{f(\vec{i},\vec{j}\,)}>\frac{2^m\sqrt{3}-1}{2^m-1}$, where $f(\vec{i},\vec{j}\,)$ is defined in Eq.(E5). From Lemmas 4 and 5, we get $v^*<\frac{2^m-1}{2^m\sqrt{3}-1}\leq v'$. Hence, there is a nonzero measure of pure states whose Werner states cannot be completely detected by homogenous linear Bell inequalities.

Case 2. General linear Bell inequalities. From Lemma 2 and Eq.(G2), we have $v'\geq \frac{2^m-1}{2^m(\sqrt{3}\sum_{i=1}^{m-1}\gamma_i^{-1}+1)-1}$ when $\|{\cal B}\|_c \geq \gamma_i\|{\cal B}_i\|_c$, $i=1, \cdots, m-1$. Moreover, when all $\gamma_i$s satisfy $\sum_{i=1}^{m-1}\frac{1}{\gamma_i}<\frac{1}{\sqrt{3}}{\rm Poly}(m)-1$ we get
\begin{align*}
\frac{2^m-1}{2^m(\sqrt{3}\sum_{i=1}^{m-1}\frac{1}{\gamma_i}+1)-1}>
\frac{2^m-1}{2^{m}{\rm Poly}(m)-1}> v^*
\tag{G3}
\end{align*}
from Lemmas 4 and 5. It means that there is a nonzero measure of pure states whose Werner states cannot be completely detected by a general linear Bell inequality if the corresponding Bell operators ${\cal B}$ and ${\cal B}_i$ satisfy $\|{\cal B}\|_c \geq \gamma_i \|{\cal B}_i\|_c$ for $i=1, \cdots, m-1$, where $\gamma_i$ satisfy $\sum_{i=1}^{m-1}\frac{1}{\gamma_i}<\frac{1}{\sqrt{3}}{\rm Poly}(m)-1$. Furthermore, for $m\leq 6$, from Table II there is a nonzero measure of pure states whose Werner states cannot be detected by general linear Bell inequalities.

\section*{Appendix H. Examples}

{\bf Example 1}. CHSH inequality \cite{CHSH}, Mermin inequality \cite{Mermin} and Svetlichny inequality \cite{Sv,BGP,CWKO} are homogeneous Bell inequalities that cannot detect Werner states of almost all generalized GHZ states or Werner states of most general pure states.

{\bf Example 2}. CH inequality \cite{CH} depending on probabilities is given by
\begin{align*}
r_{1}+r'_0+r_0r'_1-r_1r'_1-r_1r'_0-r_0r'_0\leq 1
\tag{H1}
\end{align*}
for $r_i,r'_i\in [0,1]$. It is equivalent to the following operator inequality
\begin{align*}
\langle \textbf{A}_{1}+\textbf{B}_0+\textbf{A}_0\otimes\textbf{B}_1
-\textbf{A}_1\otimes\textbf{B}_1-\textbf{A}_1\otimes\textbf{B}_0-\textbf{A}_0\otimes\textbf{B}_0\rangle_c\leq 4
\tag{H2}
\end{align*}
for $\|\textbf{A}_i\|,\|\textbf{B}_i\|\leq 1$. Moreover, we easily obtain that $\gamma_1=\frac{4}{3}$ from $\langle \textbf{A}_{1}+\textbf{B}_0+\textbf{A}_0\otimes\textbf{B}_1 -\textbf{A}_1\otimes\textbf{B}_1-\textbf{A}_1\otimes\textbf{B}_0-\textbf{A}_0\otimes\textbf{B}_0\rangle_c \leq 3$ and $\gamma_2=4$ from $\langle \textbf{B}_{0}\rangle\leq 1$. From Theorems 1 and 2, CH inequality cannot detect Werner states of the most generalized GHZ states or Werner states of general pure states with nonzero measure.

{\bf Example 3}. Scarani-Acin-Schenck-Aspelmeyer inequality
\begin{align*}
\langle \textbf{A}_0\otimes\textbf{C}_1\otimes\textbf{D}_0+\textbf{A}_0\otimes\textbf{C}_0\otimes\textbf{D}_1
+\textbf{A}_1\otimes\textbf{B}_0\otimes\textbf{C}_0\otimes\textbf{D}_0
-\textbf{A}_1\otimes\textbf{B}_0\otimes\textbf{C}_1\otimes\textbf{D}_1\rangle_c\leq 2
\tag{H3}
\end{align*}
has been used for detecting four-qubit cluster state \cite{SASA}. Note that all correlations include $A_i$.  This inequality is equivalent to a homogeneous Bell inequality as Example 1 using Lemma 3.

{\bf Example 4}. Brunner-Sharam-Vertesi inequalities
\begin{align*}
I_4=[-1 \,\,\,\, -1; -2\,\,\,\, 0\,\,\,\, - 2;\,\,\,\, -2\,\,\,\, 1\,\,\,\, 1\,\,\,\, -2]\leq  8
\tag{H4}
\end{align*}
for $4$-partite entanglement
and
\begin{align*}
I_5=[0\,\,\,\, 0 ; -2\,\,\,\, 0\,\,\,\, - 1 ; 0\,\,\,\, 0\,\,\,\, 0\,\,\,\, 0 ; -4\,\,\,\, 0 \,\,\,\,2\,\,\,\, 0\,\,\,\, 1 ]\leq 15
\tag{H5}
\end{align*}
for $5$-partite entanglement as general Bell inequalities including sub-correlations have been used to test the structure of multipartite entanglement \cite{BSV}, where we have taken use of the symmetric notation \cite{BSV}. For the inequality presented in Eq.(H4), we can obtain $\gamma_1=\frac{1}{2}$, $\gamma_2=\frac{4}{5}$, $\gamma_3=\frac{4}{3}$ and $\gamma_4=4$ assisted by computer. If we take use of Bell inequality $|I_4|\leq 32$, we have larger $\gamma_i$ as $\gamma_1=\frac{8}{5}$, $\gamma_2=\frac{16}{5}$, $\gamma_3=\frac{16}{3}$ and $\gamma_4=16$, which satisfy the lower bounds presented in Table II. From Theorems 1 and 2, there is a nonzero measure of generalized GHZ states or general pure states whose Werner states cannot be detected by this inequality. Moreover, for the inequality shown in Eq.(H5), we obtain $\gamma_1=\frac{15}{29}$, $\gamma_2=\frac{5}{4}$ and $\gamma_3=\frac{15}{4}$. Similarly, for Bell inequality $|I_5|\leq 45$ we have larger $\gamma_i$ as $\gamma_1=\frac{45}{29}$, $\gamma_2=\frac{15}{4}$ and $\gamma_3=\frac{45}{4}$ that satisfy the lower bounds shown in Table II. Hence, from Theorems 1 and 2 there is a nonzero measure of generalized GHZ states or general pure states whose Werner states cannot be detected by this inequality.

\end{document}